\RequirePackage{fix-cm}
\documentclass[aps,english,twocolumn,prd,superscriptaddress, nofootinbib]{revtex4-2}

\usepackage[T1]{fontenc} \usepackage[utf8]{inputenc} \usepackage{babel}

\usepackage{mathtools} \usepackage{amsfonts} \usepackage{mathrsfs}
\usepackage{bbm} \usepackage{slashed} \usepackage{nicefrac}

\usepackage{graphicx} \usepackage{color} \usepackage{array}

\usepackage{placeins} \usepackage{booktabs} \usepackage{makecell} %
\usepackage{floatrow}

\usepackage{xspace} \usepackage{siunitx} \usepackage{hyperref}
\usepackage[nameinlink]{cleveref} \usepackage{appendix}

\usepackage{xifthen} \usepackage{xcolor} \hypersetup{ colorlinks,
	linkcolor={red!75!black}, citecolor={blue!75!black}, urlcolor={blue!75!black} }

\def\eq#1{\labelcref{#1}}

\graphicspath{{./figures/}}

\newcommand{\iu}{{i\mkern1mu}}

\colorlet{darkred}{red!50!black} \colorlet{darkgreen}{green!60!black}

\hypersetup{
	pdftitle={Towards sampling complex actions},
	pdfauthor={Kades,G\"arttner,Gasenzer,Pawlowski},
	pdfkeywords={sign problem}
	{complex langevin} {monte carlo} {lattice physics}
	{machine learning},
	bookmarksopen=true,
	bookmarksopenlevel=2,
	bookmarksnumbered=true
}

\makeatletter
\let\cat@comma@active\@empty
\makeatletter

\begin{document}
	
	\title{Towards sampling complex actions}
	
	\author{Lukas Kades} \affiliation{Institut f\"ur Theoretische Physik, Ruprecht-Karls-Universit\"at
		Heidelberg, Philosophenweg 16, 69120 Heidelberg, Germany}
	
	\author{Martin G\"arttner} \affiliation{Physikalisches Institut, Universit\"at Heidelberg, Im Neuenheimer Feld 226, 69120 Heidelberg, Germany} \affiliation{Kirchhoff-Institut f\"ur Physik,
		Ruprecht-Karls-Universit\"at Heidelberg, Im Neuenheimer Feld 227, 69120
		Heidelberg, Germany} \affiliation{Institut f\"ur Theoretische Physik, Ruprecht-Karls-Universit\"at
		Heidelberg, Philosophenweg 16, 69120 Heidelberg, Germany}
	
	\author{Thomas Gasenzer} \affiliation{Kirchhoff-Institut f\"ur Physik,
		Ruprecht-Karls-Universit\"at Heidelberg, Im Neuenheimer Feld 227, 69120
		Heidelberg, Germany} \affiliation{Institut f\"ur Theoretische Physik,
		Ruprecht-Karls-Universit\"at Heidelberg, Philosophenweg 16, 69120 Heidelberg,
		Germany} \affiliation{ExtreMe Matter Institute EMMI, GSI, Planckstra{\ss}e 1,
		64291 Darmstadt, Germany}
	
	\author{Jan M. Pawlowski} \affiliation{Institut f\"ur Theoretische Physik,
		Ruprecht-Karls-Universit\"at Heidelberg, Philosophenweg 16, 69120 Heidelberg,
		Germany} \affiliation{ExtreMe Matter Institute EMMI, GSI, Planckstra{\ss}e 1,
		64291 Darmstadt, Germany}
	
	
	\begin{abstract}
		Path integrals with complex actions are encountered for many physical systems ranging from spin- or mass-imbalanced atomic gases and graphene to quantum chromo-dynamics at finite density to the non-equilibrium evolution of quantum systems. Many computational approaches have been developed for tackling the sign problem emerging for complex actions. Among these, complex Langevin dynamics has the appeal of general applicability. One of its key challenges is the potential convergence of the dynamics to unphysical fixed points. The statistical sampling process at such a fixed point is not based on the physical action and hence leads to wrong predictions. Moreover, its unphysical nature is hard to detect due to the implicit nature of the process. In the present work we set up a general approach based on a Markov chain Monte Carlo scheme in an extended state space. In this approach we derive an explicit real sampling process for generalized complex Langevin dynamics. Subject to a set of constraints, this sampling process is the physical one. These constraints originate from the detailed-balance equations satisfied by the Monte Carlo scheme. This allows us to re-derive complex Langevin dynamics from a new perspective and establishes a framework for the explicit construction of new sampling schemes for complex actions.
	\end{abstract}

	\maketitle
	
	
	\section{Introduction}
	\label{sec:intro}
	
	The quantum statistical properties of a physical system are described by its partition function $Z$. In particular, observables and correlation functions can be computed according to the path integral
	\begin{equation}
	\label{eq:ExpectationValue}
	\langle \mathcal{O}(\phi) \rangle = \frac{1}{Z}\int \mathcal{D}\phi\, \mathcal{O}(\phi) \exp(-S(\phi))\,,
	\end{equation}
	where $S(\phi)$ denotes the Euclidean action of the considered system. The field $\phi$ describes the state of the system in Euclidean spacetime. In lattice physics, spacetime is discretized and the fields live on a d+1-dimensional hybercubic lattice~\cite{Batrouni1985, Damgaard1987}.
			
	If $S(\phi)$ is real-valued, the weight $Z^{-1}\exp(-S(\phi))$ can be interpreted as a probability measure. This analogy enables a numerical computation of correlation functions based on standard Monte Carlo techniques. Therefore, the computation of a Euclidean quantum field theory turns into a simulation of a statistical system which is coupled to a heat bath. Its properties can be accessed by computing expectation values of a stationary distribution generated by a stochastic process in some fictitious time. This approach is referred to as stochastic quantization~\cite{Parisi1981, Klauder1983, Damgaard1987, Namiki1992}.
	
	In many physical theories, the measure $\exp(-S(\phi))$ turns out to be complex. Besides real-time dynamics~\cite{Berges2007, Cohen2015, Alexandru2016, Kanwar2021}, this is, for example, the case for the Hubbard model~\cite{Loh1990, Scalapino2007, LeBlanc2015, Ulybyshev2019, Ulybyshev2020, Berger2020}, for spin- or mass-imbalanced systems~\cite{Braun2013, Gubbels2013, Rammelmueller2017, Alexandru20182, Rammelmueller2020} and graphene~\cite{Castro2009, Ulybyshev2013, Smith2014} or for quantum chromo-dynamics at finite density~\cite{Hasenfratz1992, Muroya2003, Stephanov2006, Aarts2008, Forcrand2010, Seiler2013, Sexty2014, Mori2018, Joseph2019, Kashiwa2019, Alexandru2020, Attanasio2020}. In these theories, the fermionic part of the system contributes a multiplicative fermion determinant to the path integral measure in Eq.~\eq{eq:ExpectationValue}. The determinant can be complex, resulting in an oscillating behaviour of the integrand. This makes a direct application of stochastic techniques infeasible since the integral weight no longer represents a probability measure. Due to a possible cancellation of negative and positive contributions in the integral, almost every configuration is equally important. As a result, configurations with a small or negative Boltzmann weight are as important as samples with a large weight. To get numerical results with small errors the entire configuration space needs to be covered by the simulation method, which is infeasible for high-dimensional systems. This limitation is referred to as the \textit{sign problem}~\cite{Klauder1983, Parisi1984, Troyer2005, Berger2020, Alexandru2020}.
		
	Complex Langevin dynamics is considered as a promising numerical method for computing observables for systems that are subject to a sign problem~\cite{Klauder1983, Parisi1984}. However, two major problems of the method are numerical instabilities, such as, runaway trajectories, and a possible convergence to an unphysical solution~\cite{Ambjorn1985, Aarts2008, Berger2020, Aarts20102, Attanasio2019, Aarts2012}. 
	
	Applying complex Langevin to models plagued by a sign problem is an active area of research, see~\cite{Seiler, Attanasio2020} for recent reviews. In~\cite{Alexandru2020, Berger2020} an overview of other methods tackling the sign problem is given, such as reweighting or a deformation of the integration contour into the complex plane.
	
	Here we introduce a framework that generalizes complex Langevin dynamics and allows deriving the algorithm from first principles. The framework comprises an interpretation of complex Langevin dynamics as a standard Markov chain Monte Carlo algorithm. This point of view opens up perspectives on making use of knowledge from several decades in research on Monte Carlo algorithms. Beyond providing a foundation for complex Langevin dynamics, the framework serves as a basis for deriving new algorithms for theories with and without a sign problem. We open up a perspective that has the potential to facilitate the development and improvement of novel algorithms and to better evaluate and understand existing approaches to tackling the sign problem.
		
	A brief recapitulation on stochastic quantization and complex Langevin dynamics is provided in Chapter~\ref{sec:RelatedWork} to make the manuscript self-contained. Chapter~\ref{sec:MainContributions} relates complex Langevin dynamics with first principles of our framework for computing observables of problems with a sign problem. We continue in Chapter~\ref{sec:MarkovChainMonteCarloSamplingInAuxiliaryDimensions} with a reminder on Markov chain Monte Carlo methods making use of auxiliary dimensions and point out important differences to our framework. \textit{Substitution Sampling} is introduced in Chapter~\ref{sec:SubstituionSampling} as a Markov chain Monte Carlo method that allows the numerical computation of observables based on the provided formal framework of the two previous chapters. Examples for substitution sampling algorithms with similar properties to complex Langevin dynamics are given in Chapter~\ref{sec:FurtherLangevinLikeAlgorithms}. A further, different kind of substitution sampling algorithm is defined in Chapter~\ref{sec:SHMCS}. The derived algorithms and the formal framework are numerically benchmarked in Chapter~\ref{sec:NumericalResults}. The work ends with a conclusion and an
	outlook in Chapter~\ref{sec:ConclusionAndOutlook}.
	
	\section{Stochastic quantization and complex Langevin dynamics}
	\label{sec:RelatedWork}
	
	The objective of this chapter is to introduce the basics of stochastic quantization and, specifically, complex Langevin dynamics.
	
	\subsection{A toy model}
	
	To illustrate how the sign problem appears in this context, we choose a toy model, which we will later use for comparison of different algorithms. The zero-dimensional polynomial model~\cite{Ambjorn1985, Aarts2013, Nagata2016} is defined by the action
	\begin{equation}
	\label{eq:PolynomialModel}
	S(\phi) = \frac{1}{2}\left(\sigma_{\text{Re}} + \iu \sigma_{\text{Im}}\right)\phi^2 + \frac{\lambda}{4} \phi^4\,,
	\end{equation}
	a function depending on the real-valued scalar field $\phi\in\mathbb{R}$ and real-valued couplings $(\lambda, \sigma_\text{Re}, \sigma_\text{Im})$\footnote{This toy model is widely used in studying the sign problem. It is one of the simplest non-trivial quantum mechanical models~\cite{Ambjorn1985} and describes, e.g., a single-mode relativistic interacting Bose gas at nonzero chemical potential $\mu \propto \sigma_{\text{Im}}$~\cite{Aarts2009, Aarts2013}.}.
	
	The objective is to compute observables over $\phi$:
	\begin{equation}
	\label{eq:Observable}
	\langle \mathcal{O}(\phi) \rangle = \int_{-\infty}^{\infty} \text{d}\phi\,\mathcal{O}(\phi)\rho(\phi)\,,
	\end{equation}
	with equilibrium Boltzmann measure
	\begin{equation}
	\label{eq:BoltzmannDistribution}
	\rho(\phi)=\frac{1}{Z}\exp\left(-S(\phi)\right)\,,
	\end{equation}
	which is normalized by the partition function
	\begin{equation}
	Z = \int_{-\infty}^{\infty} \text{d}\phi\,\exp(-S(\phi))\,,
	\end{equation}
	and where the usual energy divided by temperature, $\beta E$, is upgraded, in quantum theory, to the Euclidean action $S$.
	
	Standard Monte Carlo methods rely on sampling from a probability distribution. Since the action~\eq{eq:PolynomialModel} is complex, these methods are, at first sight, inapplicable here.
	
	\subsection{Real Langevin dynamics} \label{sec:AppendixComplexLangevinDynamics}
	
	The Langevin equation, originally formulated to model Brownian motion~\cite{Langevin1908}, is central to the stochastic quantization approach to quantum field theory~\cite{Parisi1981, Klauder1983, Damgaard1987, Namiki1992}. 
	In the simplest case, it describes the evolution of a real, scalar field $\phi(x)$, governed by a real Euclidean action $S(\phi)$, in
	an additional, fictitious time dimension, the Langevin time $\tau$. It reads
	\begin{equation}
	\label{eq:LangevinDynamics}
	\frac{\partial}{\partial \tau}
	\phi(\tau) = -\frac{\delta S}{\delta \phi(\tau)} + \eta(\tau)\,,
	\end{equation}
	where we suppress the dependence of $\phi$ and $\eta$ on $x$ for brevity.
	
	Similar to thermal fluctuations in a thermodynamic system with
	energy $E$, the noise term $\eta$ emulates quantum fluctuations in the
	case of a Euclidean quantum field theory. The distribution of the noise term $\eta$ is usually taken to be centred at zero,
	\begin{equation}
	\label{eq:GaussianNoise}
	\langle \eta(\tau)\, \eta(\tau')\rangle_\eta = 2\delta(\tau' -
	\tau)\,,\quad\langle \eta(\tau) \rangle_\eta = 0\,, \end{equation}
	where $\langle \cdot \rangle_\eta$ denotes the expectation value with respect to the noise distribution.
	A common choice for this distribution is Gaussian white noise. Under these
	conditions, the $\tau$-dependent distribution of $\phi$ is subject to the Fokker-Planck equation~\cite{Berger2020, Damgaard1987, Parisi1981},
	\begin{equation}
	\label{eq:FPE}
	\frac{\partial \rho(\phi, \tau)}{\partial \tau} = \int \textrm{d}^d x
	\frac{\delta}{\delta \phi(\tau) } \left(\frac{\delta S}{\delta \phi(\tau)} +
	\frac{\delta}{\delta \phi(\tau)}\right) \rho(\phi, \tau)\,.
	\end{equation}
	Its stationary solution is the Boltzmann distribution~\eq{eq:BoltzmannDistribution},
	\begin{equation}
	\lim\limits_{\tau\to \infty} \rho(\phi, \tau)=\rho(\phi)\,.
	\end{equation}
	For the case of a real-valued action $S(\phi)$ considered here, it can be shown that the Langevin evolution converges, in the
	limit $\tau \rightarrow \infty$, to the desired equilibrium distribution as a stationary solution and that the convergence is exponentially
	fast \cite{Damgaard1987}. After an equilibration period of time $\bar{\tau}$, observables can be evaluated by
	\begin{equation}
	\langle \mathcal{O}(\phi)\rangle_\rho \simeq \frac{1}{T} \int_{\bar{\tau}}^{\bar{\tau} + T} \text{d}\tau\,\mathcal{O}(\phi(\tau))\,,
	\end{equation}
	where $T$ is a suitable time to correctly estimate equilibrium expectation values through temporal averaging. Hence, by discretizing both $x$ and $\tau$, the
	Langevin equation provides a means to sample lattice quantum field theories, as
	long as the accumulation of numerical errors caused by the discretization is
	controllable.
	
	\subsection{Complex Langevin dynamics}
	\label{sec:ComplexLangevinDynamics}
	
	A generalization of stochastic quantization to complex distributions $\rho(\phi)$ has been proposed as a means to numerically access observables of systems with a sign problem~\cite{Klauder1983, Parisi1984, Damgaard1987}. For a complex action $S(\phi)$, the Langevin equation~\eq{eq:LangevinDynamics} in general describes an evolution leading to complex values for $\phi=\phi_x + \iu \phi_y$. The real and imaginary components are then commonly evolved according to the equations
	\begin{align}
	\label{eq:ComplexLangevin} \frac{\partial}{\partial \tau}
	\phi_x(\tau) &= -\text{Re}\left[\frac{\delta S}{\delta \phi(\tau)}\bigg|_{\phi_x
		+ \iu \phi_y}\right] + \eta_x(\tau)\,,\nonumber\\[1ex] \frac{\partial}{\partial \tau}
	\phi_y(\tau) &= -\text{Im}\left[\frac{\delta S}{\delta \phi(\tau)}\bigg|_{\phi_x
		+ \iu \phi_y}\right]\,,
	\end{align}
	where only the equation for $\phi_x$ is of the Langevin form with, commonly, white noise $\eta_x$, while the equation for $\phi_y$ describes a pure drift. A noise term in the imaginary part can introduce stability problems which is why the evolution is usually driven by purely real noise~\cite{Aarts2010, Aarts2013}. We will later show that the missing imaginary noise term is also justified on formal grounds.
	
	The stochastic process converges to solutions governed by a real-valued steady-state distribution $P(\phi_x, \phi_y) = \lim\limits_{\tau \to \infty} P(\phi_x, \phi_y, \tau)$ in the $\phi_x$-$\phi_y$-plane. Observables
	\begin{align}
	\langle \mathcal{O}(\phi_x& + \iu \phi_y) \rangle_{p} \nonumber\\[1ex]&= \int \text{d}\phi_x \int \text{d}\phi_y\,\mathcal{O}(\phi_x + \iu \phi_y) P(\phi_x, \phi_y, \tau)\,
	\end{align}
	can be numerically computed by sampling from the resulting distribution. The expectation values coincide under certain constraints with the expectation values
	with respect to the original complex distribution $\rho(\phi, \tau)$,
	\begin{equation}
	\label{eq:ConvergenceCLE}
	\langle \mathcal{O}(\phi_x + \iu\phi_y) \rangle_{P} = \langle \mathcal{O}(\phi) \rangle_{\rho}\,.
	\end{equation}
	In the standard approach to analysing the convergence of complex Langevin, one compares two independent time-dependent stochastic processes~\eq{eq:ComplexLangevin}, namely, the evolutions of the distribution $P(\phi_x, \phi_y, \tau)$, and of the underlying complex distribution $\rho(\phi, \tau)$ by means of their respective Fokker-Planck equations~\cite{Aarts2010, Seiler, Berger2020}. Details about the existence and the properties of a stationary distribution $P(\phi_x, \phi_y)$, which satisfies Eq.~\eq{eq:ConvergenceCLE}, can be found, for example, in~\cite{Salcedo2016}.
	
	An issue with the complex Langevin ansatz is the absence of a guaranteed convergence to the correct equilibrium distribution, which, in most cases, can only be verified a posteriori. As pointed out, a correct convergence is ensured only under certain conditions that have been elaborated in the past, see, for example~\cite{Namiki1992, Salcedo2016, Seiler, Aarts2010, Aarts2011, Aarts2013, Aarts2018, Nagata2016, Nagata2018, Scherzer2019, Seiler2020}.
	
	Besides the requirement of ergodicity, model actions and distributions should ideally be holomorphic. Studying models with meromorphic poles is, in principle, also possible, but more care has to be taken to ensure convergence~\cite{Aarts2017}. Lastly, the numerically sampled distributions of the observables need to decay fast enough in the imaginary direction~\cite{Nagata2016, Nagata2018}.
	
	There exist different ways for checking if these criteria are fulfilled. One important way involves computing boundary terms by considering the derivative of a quantity $F_\mathcal{O}(t, \tau)$ with respect to the Langevin time $\tau$~\cite{Aarts2010, Aarts2011, Scherzer2019, Scherzer2020}. The function $F_\mathcal{O}(t, \tau)$ interpolates between the two observables in Eq.~\eq{eq:ConvergenceCLE}. Other approaches are based, for example, on an analysis of the decay of the sampled probability distribution~\cite{Nagata2016, Nagata2018, Nishimura2015}.
	
	\section{Summary of main results}
	\label{sec:MainContributions}
	
	In this work, we formulate a general approach for deriving Markov chain Monte Carlo algorithms for computing expectation values for theories with a sign problem. The resulting framework rests on a reformulation of the path integral~\eq{eq:ExpectationValue} as a one-dimensional\footnote{One dimensional here refers to the important fact that, even if the integration variable $\phi$ is allowed to become complex, one still integrates over $\phi$ only. This in contrast to, e.g., a coherent-state path integral $~\sim\int\text{d}\phi\,\text{d}\phi^\star\,\exp(-S)$, which, in that sense, is two dimensional.} stochastic integral in the complex plane. The approach is sketched in more detail in Sec.~\ref{sec:KeyFindings}. Implications on the numerical sampling framework, presented in this work, are outlined in Sec.~\ref{sec:Structure}. Complex Langevin dynamics represents one algorithm that can be derived within this framework.
	
	\subsection{Motivation}
	
	The standard approach in formulating complex Langevin dynamics consists of inserting the complex action into the Langevin equation and proving the validity of the solutions by a comparison with the associated Fokker-Planck equations, cf. Eq.~\eq{eq:FPE} for the case of real Langevin. Here, we take a different route and derive complex Langevin dynamics from first principles.
	
	\begin{figure}
		\includegraphics[width=\linewidth]{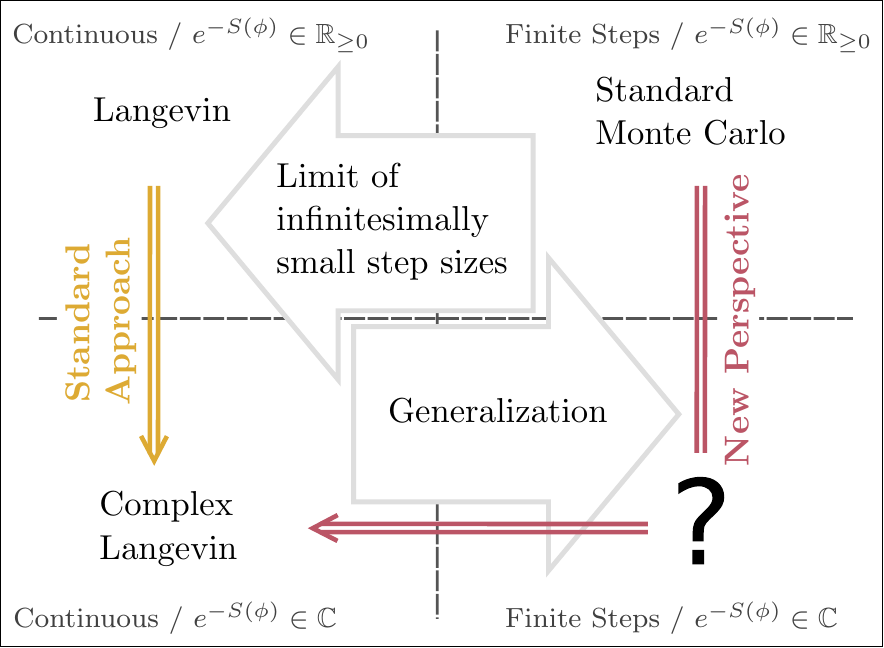}
		\caption{Comparison of the standard approach to deriving complex Langevin dynamics and of the perspective provided in this work. By taking the limit of infinitesimally small step sizes, the Markov chain turns into a continuous evolution in state space. This kind of dynamics corresponds to the left-hand side of the graphics. We pursue the goal to generalize complex Langevin dynamics as a Monte Carlo algorithm for complex measures that also works with finite step sizes in configuration space.}
		\label{fig:Perspectives}
	\end{figure}
	
	While complex Langevin dynamics is thus identified as a valid means for evaluating Eq.~\eq{eq:ExpectationValue} for complex-valued actions $S$, it nevertheless still suffers from the numerical problem of runaway processes as well as convergence to unphysical solutions~\cite{Aarts2018, Salcedo2016, Nagata2018, Nagata2016, Berger2020, Attanasio2020, Scherzer2019, Scherzer2020, Seiler2020, Nishimura2015, Bluecher2018, Alvestad2021}. Moreover, the continuous evolution of Eq.~\eq{eq:ComplexLangevin} cannot be straightforwardly applied to models of discrete-valued fields $\phi$ such as of spin systems.
	
	With the framework introduced in the following, we aim to pave the way for two long-term goals:
	
	\begin{itemize}
		\item A generalization of complex Langevin dynamics that allows developing numerically more stable sampling algorithms.
		\item A numerical computation of expectation values for discrete systems with a sign problem which does not rely on reweighting but entails sampling in an extended state space.
	\end{itemize}
	
	In Fig.~\ref{fig:Perspectives}, we relate known techniques and their derivations with the chosen path of our work to achieve these two goals.
	
	\subsection{Key insights}
	\label{sec:KeyFindings}
	
	Our central task is to evaluate expectation values of observables $\mathcal{O}(\phi)$ with respect to the complex distribution $\rho(\phi)$ defined in Eq.~\eq{eq:BoltzmannDistribution}, depending, for the first, on a real-valued field $\phi$,
	\begin{equation}
	\label{eq:ExpectationValueMC}
	\langle \mathcal{O}(\phi) \rangle_{\rho} = \int_{a}^{b} \text{d}\phi\,\mathcal{O}(\phi) \rho(\phi)\,,
	\end{equation}
	with integral boundaries $a$ and $b$.
	
	We substitute the field variable $\phi$ by
	\begin{equation}
	\label{eq:Substitution}
	\phi = \phi(\phi_x) = \phi_x + \iu \phi_y\,,\quad \text{d}\phi = \text{d}\phi_x\,,
	\end{equation}
	where the integration variable is now the real field $\phi_x$, and $\phi_y$ is, for the moment, just a constant. The integral turns into
	\begin{equation}
	\langle \mathcal{O}(\phi) \rangle_{\rho} = \int_{a - \iu \phi_y}^{b - \iu \phi_y} \text{d}\phi_x\,\mathcal{O}(\phi_x + \iu \phi_y) \rho(\phi_x + \iu \phi_y)\,.
	\end{equation}
	If the integral is invariant with respect to a shift of its boundaries by $\iu\phi_y$, we can reset the integral bounds back to $a$ and $b$, such that the integral reads
	\begin{equation}
	\label{eq:IntegralCLE}
	\langle \mathcal{O}(\phi) \rangle_{\rho} = \int_a^b \text{d}\phi_x\,\mathcal{O}(\phi_x + \iu \phi_y) \rho(\phi_x + \iu \phi_y)
	\end{equation}
	and, as a result, becomes independent of $\phi_y$. We will later identify $\phi_y$ with the imaginary part of the field in the complex Langevin evolution. Under these conditions, we can express the expectation value as the mean over multiple, arbitrary values of $\phi_y$,
	\begin{equation}
	\label{eq:NumericIntegralCLE}
	\langle \mathcal{O}(\phi) \rangle_{\rho} = \frac{1}{N}\sum_{i=1}^{N}\int_a^b \text{d}\phi_x\,\mathcal{O}(\phi_x + \iu \phi_{y; i}) \rho(\phi_x + \iu \phi_{y; i})\,,
	\end{equation}
	and it will be sufficient to assume that the above invariance holds for the range of values of $\phi_y$ appearing in this sum. We furthermore assume that there exists a numerical method for sampling $\phi_x$ from $\rho$. In this case, we can express the integral on the right-hand side as the mean value
	\begin{equation}
	\langle \mathcal{O}(\phi) \rangle_{\rho} = \frac{1}{N}\sum_{i=1}^{N} \frac{1}{M} \sum_{j=1}^{M} \mathcal{O}(\phi_{x; ij} + \iu \phi_{y; i})\,.
	\end{equation}
	Note that $\phi_x$ depends on both $i$ and $j$, meaning that one draws the samples $\lbrace \phi_{x; ij}\rbrace$ for a fixed $\phi_{y; i}$.
	
	In the last step, we argue that we can mix $\phi_{x; ij}$'s belonging to different values of $\phi_{y; i}$ as long as changes in $\phi_{y; i}$ do not introduce further correlations between the updated $\phi_y$ and the sampled $\phi_x$, as will be discussed in more detail below. This can be implemented by allowing only infinitesimally small changes in $\phi_y$, independent of the sampling probability for $\phi_x$.
	
	Under the above condition, we can keep the sampling index $i$ as the only one in the sum over, still, $MN$ samples, implying that we no longer consider separate evolutions for a fixed $\phi_{y; i}$, but smoothly mix the respective evolutions in $\phi_x$. So the index $i$ counts the combined update step of both $\phi_x$ and $\phi_y$, performed, e.g., in complex Langevin dynamics,
	\begin{equation}
	\langle \mathcal{O}(\phi) \rangle_{\rho} = \frac{1}{MN}\sum_{i=1}^{MN} \mathcal{O}(\phi_{x; i} + \iu \phi_{y; i})\,.
	\end{equation}
	This expression is eventually to be understood as a numerical expectation value determined from samples $(\phi_x, \phi_y)$ which a stochastic process generated according to the complex distribution $\rho(\phi_x + \iu \phi_y)$, i.e.:
	\begin{equation}
	\label{eq:NumExpectationValueCLE}
	\langle \mathcal{O}(\phi_x + \iu \phi_y) \rangle_{\rho} = \frac{1}{MN}\sum_{i=1}^{MN} \mathcal{O}(\phi_{x; i} + \iu \phi_{y; i})\,.
	\end{equation}
	We conclude that we can reinterpret the computation of the integral in Eq.~\eq{eq:ExpectationValueMC} as that of the expectation value of a combined process in $\phi_x$ and $\phi_y$, where it needs to be guaranteed that stochastic changes according to some transition probability take place only in the $\phi_x$ direction. This ensures that we independently compute a mean value with respect to the imaginary part of the field. As a result, the expectation value in the extended state space of complex $\phi$ still reflects the degrees of freedom of the integral~\eq{eq:ExpectationValueMC}.
	
	In summary, we can write
	\begin{equation}
	\label{eq:ExpectationValueRho}
	\langle \mathcal{O}(\phi) \rangle_{\rho} = \langle \mathcal{O}(\phi_x + \iu \phi_y) \rangle_{\rho}\,,
	\end{equation}
	as long as the invariance of the integral~\eq{eq:ExpectationValueMC} under imaginary shifts of the boundaries holds and the underlying stochastic process has a vanishing variance along the $\phi_y$ direction. Note that the expectation value on the right-hand side is considered to be computed by means of a Markov chain Monte Carlo algorithm with respect to $\rho(\phi_x + \iu \phi_y)$, but constrained by the conditions provided above.
	
	These conditions imply that the eventually obtained higher-dimensional probability distribution, denoted as $P(\phi_x, \phi_y)$, is different from $\rho(\phi_x + \iu \phi_y)$. This point of view is in strong contrast to standard considerations of complex Langevin dynamics where a Fokker-Planck equation in both $\phi_x$ and $\phi_y$ and, therefore, an expectation value with respect to a distribution $P(\phi_x, \phi_y)$ is analysed~\cite{Aarts2010, Seiler, Berger2020}. Furthermore, the restriction to an infinitesimal step size in the $\phi_y$ direction underscores findings of a higher numerical stability for vanishing imaginary noise~\cite{Aarts2010, Aarts2013}. It also corroborates the finding that a distribution $P(\phi_x, \phi_y)$ which decays sufficiently fast in the imaginary direction ensures correct convergence~\cite{Nagata2016, Nagata2018}.
	
	Due to the restrictions discussed above, we can compute the expectation value on the right-hand side of Eq.~\eq{eq:ExpectationValueRho} only by an extrapolation to a vanishing change in $\phi_y$. In practice, this can be achieved by performing multiple simulations with small step sizes and an extrapolation of the resulting observables.
	
	At first sight, evaluating an expectation value with respect to the complex distribution $\rho(\phi_x + \iu \phi_y)$ over complex fields instead of $\rho(\phi)$ over real $\phi$ does not improve a numerical sampling due to difficulties in defining real-valued transition probabilities. However, the imaginary part $\phi_y$ of the field introduces an additional degree of freedom. In contrast to that of the real part $\phi_x$, the dynamics of $\phi_y$ is only constraint by the proposed restriction to an infinitesimally small update. We point out once more that this is the reason for a different distribution $P(\phi_x, \phi_y)$ observed after sampling.
	
	Hence, it is important to understand that there is a difference between sampling from a higher-dimensional probability distribution defined in $\phi_x$ and $\phi_y$ and a sampling of the complex distribution $\rho(\phi_x + \iu \phi_y)$ subject to the above restrictions. In this work, we refer only to the latter case and discuss how to set up a Markov chain Monte Carlo algorithm with respect to this numerical sampling procedure.
	
	\subsection{Key results}
	\label{sec:Structure}
	
	We will show that, in the case of complex Langevin dynamics, the introduced additional degree of freedom can be used to render the transition probabilities of a corresponding Markov chain Monte Carlo algorithm real and positive. As a result numerical sampling from the complex distribution $\rho(\phi_x + \iu \phi_y)$ becomes possible.
	
	In the following sections,
	\begin{itemize}
		\item we set the ideas laid out in Sec.~\ref{sec:KeyFindings} on firm grounds and compare a numerical sampling of the expectation value in Eq.~\eq{eq:NumericIntegralCLE} with other algorithms (Restricted Boltzmann machine and Hamiltonian Monte Carlo algorithm) where the underlying dynamics also takes place in an extended higher-dimensional state space, cf. Sec.~\ref{sec:ComplexLangevinVersusHMCRBM};
		\item we give a reminder on general aspects of Monte Carlo sampling in higher dimensions, cf. Secs.~\ref{sec:MasterEquation} and~\ref{sec:Equilibrium};
		\item we provide a possible approach to constructing transition probabilities from first principles, which allows sampling from the equilibrium distribution subject to the constraints given above, cf. Chapter~\ref{sec:SubstituionSampling};
		\item we show that complex Langevin satisfies these first principles and derive the dynamics based on this approach, cf. Sec.~\ref{sec:ComplexLangevinAsASubstitutionSamplingAlgorithm} and App.~\ref{sec:LangevinSamplingByCompensation};
		\item we provide numerical evidence in support of our approach by use of other algorithms that are built on the same first principles as complex Langevin dynamics, for the cases of a complex action and a real action, cf. Chapters~\ref{sec:FurtherLangevinLikeAlgorithms} and~\ref{sec:SHMCS} as well as App.~\ref{sec:CLELikeAlgorithms}.
	\end{itemize}
	
	Our work provides a framework for deriving Markov chain Monte Carlo algorithms that are, in principle, suitable to sample from a distribution given in terms of a complex action. We expect the framework to be useful for developing new algorithms for theories with a sign problem.
		
	\section{Markov chain Monte Carlo sampling in auxiliary dimensions}
	\label{sec:MarkovChainMonteCarloSamplingInAuxiliaryDimensions}
	
	In this chapter, we introduce a formal framework for developing Monte Carlo sampling algorithms on state spaces including additional auxiliary dimensions. By this we mean algorithms that work in a higher-dimensional representation space while sampling from a lower dimensional probability distribution. In Sec.~\ref{sec:ComplexLangevinVersusHMCRBM}, we generalize and embed the perspective on computing observables given in the previous chapter into this framework.
		
	\subsection{Extended state space}
	
	\begin{figure*}
		\centering \includegraphics[width=\linewidth]{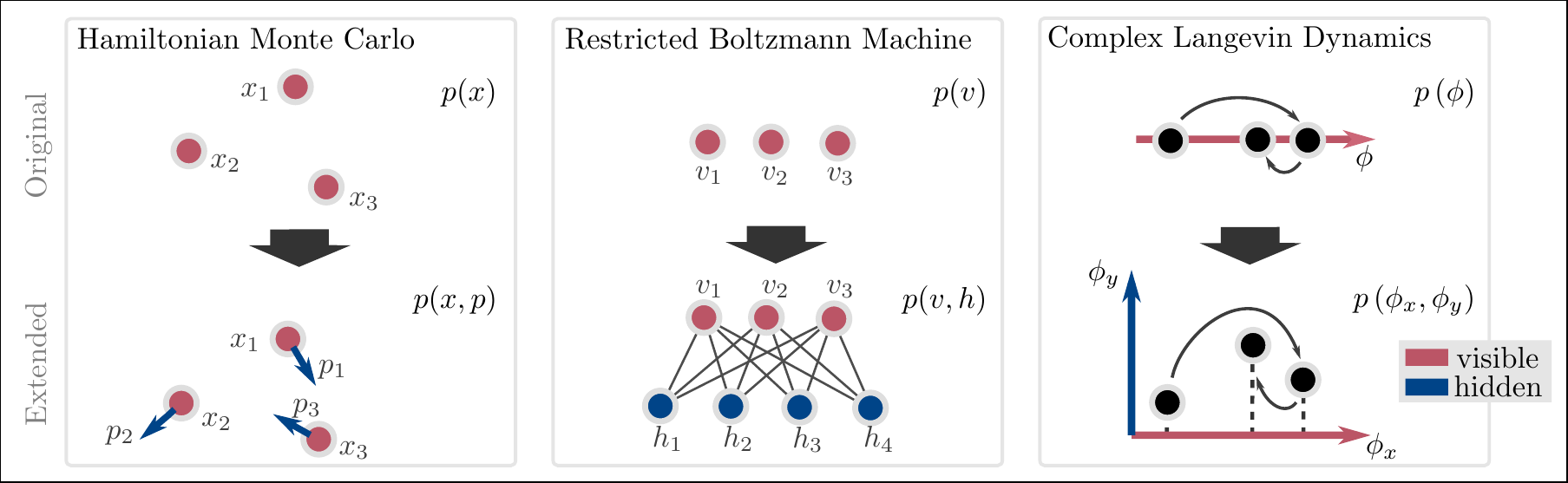}
		\caption{Comparison of different algorithms that all make use of the introduction of additional hidden variables / auxiliary dimensions. The respective Markov chain is realized in a new set of visible and hidden variables $(v, w)$. The target distribution of the different algorithms is indicated in the upper row. In the lower row, the newly introduced hidden variables are marked in blue and the visible variables in red. The probability distributions are functions of this new set of variables. In the case of complex Langevin dynamics, the field $\phi$ is promoted to a complex field where the visible variable is given by the real part $\phi_x$ and the hidden variable by the imaginary part $\phi_y$. Accordingly, the field $\phi$ is parametrized by $\phi_x + \iu \phi_y$. Observables in the original set of variables can be obtained for the Hamiltonian Monte Carlo algorithm and the restricted Boltzmann machine by marginalizing the higher-dimensional distributions. This is different for complex Langevin dynamics, where observables are expressed according to Eq.~\eq{eq:NumExpectationValueGeneral} in terms of the hidden and visible variables. Details on the algorithms can be found in different sections of this work and in the appendix.} \label{fig:SamplingInHigherDimensions}
	\end{figure*}
	
	Examples of algorithms, in which the dimension of the state space is extended by additional auxiliary dimensions include the Hamiltonian Monte Carlo algorithm (HMC) ~\cite{Duane1987}, introducing momenta for each state, or the restricted Boltzmann machine (RBM)~\cite{Smolensky1986}, with a distinction between visible and hidden layers of neurons. Both algorithms are recapitulated in App.~\ref{sec:DetailedBalanceEquationInMultipleVariablesForDifferentAlgorithms}. In the case of the Hamiltonian Monte Carlo algorithm, the extra dimensions lead to a faster exploration of the original state space. For the restricted Boltzmann machine, the introduced hidden layers are essential for the representation of a larger class of in general non-Gaussian probability distributions.
	
	Complex Langevin dynamics can also be attributed to this class of algorithms. The state space is complexified and the imaginary part represents an auxiliary variable, cf. Sec.~\ref{sec:KeyFindings}.
	
	Inspired by the RBM, we introduce auxiliary dimensions by distinguishing \textit{visible} state variables $v$ and \textit{hidden} state variables $w$. For RBMs, the visible variables are given by the neuron states in the visible layer and the hidden variables by the ones in the hidden layer. Fig.~\ref{fig:SamplingInHigherDimensions} depicts the distinction of visible (red) and hidden (blue) variables for the different algorithms. In the case of the HMC algorithm the variables $x$ are considered as visible and the momenta $p$ as hidden variables. For complex Langevin dynamics, the higher-dimensional representation is given by the complex field. The real part of the field is identified as the visible and the imaginary part as the hidden variable.
	
	The distinction between visible and hidden variables can be formally understood as follows: The original distribution $\rho(x)$ is defined over a set of variables $x$. Therefore, expectation values need to be computed by integrating over $x$. The visible variables encode the probabilistic nature of this set of original variables. Accordingly, $v$ has the same dimension as $x$, and the subspace, spanned by the visible variables, reflects the degrees of freedom of the original state space. The hidden state variables are used to improve the sampling procedure itself. The diversity of the discussed algorithms demonstrates the flexibility that originates from the introduction of additional auxiliary dimensions.
	
	\subsection{Master equation}
	\label{sec:MasterEquation}
	
	The time evolution of the distribution $\rho(x, \tau)$ of a stochastic state variable $x$, subject to transition probabilities $W(x\to x')$, is in general described by a master equation~\cite{Newman1991}:
	
	\begin{align}
	\label{eq:masterequation} \frac{\text{d}\rho(x,\tau)}{\text{d}\tau}&\nonumber\\ =&\sum_{x'} \left[\rho(x',\tau)\, W(x'\rightarrow x) - \rho(x,\tau)\, W(x\rightarrow x')\right]\,.
	\end{align}
	The right-hand side of the equation contains gain and loss terms
	for the state $x$ to go over to $x'$ and vice versa.
	
	A master equation can be formulated for the set of visible and hidden variables introduced in the previous section in the same manner as for $x$:
	\begin{align}
	\label{eq:masterequationhigherdimensions} \frac{\text{d}p(v, w,\tau)}{\text{d}\tau}=&\sum_{v', w'} \bigg[p(v', w',\tau)\,W(v', w'\rightarrow v, w) \nonumber\\[1ex] &\quad\;- p(v, w,\tau)\, W(v, w\rightarrow v', w')\bigg]\,.
	\end{align}
	In contrast to the original state $x$, the evolution is governed by transition probabilities $W(v, w \to v', w')$. They determine how the probability distribution $p(v, w,\tau)$, defined over the higher-dimensional representation space, evolves in time.
		
	\subsection{Equilibrium}
	\label{sec:Equilibrium}
	
	The standard work flow for setting up Markov chain Monte Carlo (MCMC) algorithms is to choose transition probabilities in such a way that the evolution converges, in the infinite-time limit, to an equilibrium distribution. The equilibrium distribution is expected to coincide with the probability distribution of interest. In our case, we aim at sampling from explicitly given distributions $\rho(x)$ and $p(v, w)$.
	
	The system is defined to be in equilibrium if its state distribution does not change anymore over time. This is the case when the sum on the right-hand side of the master equation~\eq{eq:masterequation} evaluates to zero. This translates into the equilibrium condition
	\begin{equation}
	\label{eq:EquilibriumConstraintOriginal}
	\rho(x',\tau) \stackrel{!}{=} \sum_{x} \rho(x,\tau)\, W(x\rightarrow x')\,,
	\end{equation}
	as can be derived by using the normalization of $W(x \to x')$ in the second term on the right-hand side of Eq.~\eq{eq:masterequation} and a respective renaming of $x$ and $x'$. In the higher-dimensional state space, it reads
	\begin{equation}
	\label{eq:EquilibriumConstraint}
	p(v', w',\tau)\stackrel{!}{=} \sum_{v, w} p(v, w,\tau)\, W(v, w\rightarrow v', w')\,.
	\end{equation}
	In App.~\ref{sec:RestrictedBoltzmannMachine}, we provide an example of how this relation is fulfilled by the equilibrium distribution of the restricted Boltzmann machine. However, it needs to be taken into account that the above condition does not guarantee a correct sampling from the desired distribution due to possible limit cycles occurring in the Markov chain~\cite{Newman1991}.
	
	A more restrictive equilibrium condition is that the transition probabilities
	satisfy the detailed-balance equation:
	\begin{equation}
	\label{eq:DetailedBalance}
	\rho(x)W(x\to x') = \rho(x')W(x'\to x)\,,
	\end{equation}
	or, in higher dimensions:
	\begin{equation}
	\label{eq:DetailedBalanceHigherDimensions}
	p(v, w) W(v, w \to v', w') = p(v', w')W(v', w'\to v, w)\,.
	\end{equation}
	Detailed balance implies that the sum on the right-hand side of the master
	equation~\eq{eq:masterequation} vanishes separately for every summand and that the process thus samples from the equilibrium distribution. The Hamiltonian Monte Carlo algorithm is discussed as an example for this approach in App.~\ref{sec:HamiltonianMonteCarlo}.
	
	The transition probabilities introduced above are used in a Markov chain to draw samples from the equilibrium distribution. Observables, as defined in Eq.~\eq{eq:ExpectationValue}, are then numerically accessible by computing expectation values according to:
	\begin{equation}
	\langle \mathcal{O}(x) \rangle = \frac{1}{N} \sum_{i=1}^{N} O(x_i)\,,
	\end{equation}
	where the sum runs over the drawn samples.
	
	Besides a time-independent state distribution, it is important that further necessary conditions, like ergodicity, are fulfilled, for more details see, for example,~\cite{Newman1991}.
		
	\subsection{Complex Langevin versus HMC / RBM}
	\label{sec:ComplexLangevinVersusHMCRBM}
	
	At this point, it is interesting to have a closer look at the use of auxiliary dimensions in the different algorithms in more detail. We will, in particular, point out differences between complex Langevin dynamics, the Hamiltonian Monte Carlo algorithm, and the restricted Boltzmann machine.
	
	For the latter two algorithms, the visible state $v$ can be identified with the state $x$ in the originally considered problem. This is an important property since it allows the numerical computation of observables in $x$ by considering just the visible states $v$. Therefore, $v=x$ and thus
	\begin{equation}
	\langle \mathcal{O}(x) \rangle_{\rho} = \langle \mathcal {O}(v) \rangle_{p} = \frac{1}{N}\sum_{i}^N \mathcal{O}(v_i)\,.
	\end{equation}
	The auxiliary, hidden variables can be ignored for the computation of observables. The mathematical argument behind this is a possible marginalization of the joint probability distribution $p(x, w)$ according to
	\begin{equation}
	\rho(x) = \int \text{d}w \, p(x, w)\,.
	\end{equation}
	This is, however, different for complex Langevin dynamics, which we show by generalizing the way expectation values are computed for this kind of dynamics.
	
	Following the line of arguments in Sec.~\ref{sec:KeyFindings}, the visible states $v$ are no longer identified with the original state $x$, but are related to them through a linear shift by the hidden variable, cf. Eq.~\eq{eq:Substitution}. The original integral is effectively computed for different substitutions $v\to v + w_i$ in terms of the hidden state variables,
	\begin{equation}
	\label{eq:SamplingGeneral}
	\langle \mathcal{O}(x) \rangle_{\rho} = \frac{1}{N}\sum_{i=1}^{N} \int_{a}^{b}\text{d}v\, \mathcal{O}(v + w_i) \rho(v + w_i)\,,
	\end{equation}
	where it is assumed that the integral is invariant under the linear shifts of the integral bounds by $-w_i$. This allows using the same integral bounds $a$ and $b$ for all different values of the hidden variables $w_i$.
	
	As a result, the auxiliary variables contribute to the numerical computation of observables, in which it is summed over samples $v_i$ instead of the continuous integrals,
	\begin{equation}
	\label{eq:NumExpectationValueGeneral}
	\langle \mathcal{O}(x) \rangle_{\rho} = \langle \mathcal{O}(v, w) \rangle_{\rho} = \frac{1}{MN}\sum_{i=1}^{MN} \mathcal{O}(v_{i}, w_{i})\,.
	\end{equation}
	In the case of complex Langevin dynamics, one may identify $v_i$ with $\phi_{x;i}$ and $w_i$ with $\iu \phi_{y;i}$, cf. Eq.~\eq{eq:NumExpectationValueCLE}.
	
	This a valid approach since we demand that the hidden variables $w$ do not undergo any stochastic evolution. In the case of complex Langevin dynamics, this is realized by a missing noise term and an extrapolation to a vanishing step size in the direction of the hidden states. Therefore, Eq.~\eq{eq:NumExpectationValueGeneral} does not compute the expectation value of a joint distribution of both the visible and the hidden states,
	\begin{equation}
	\langle \mathcal{O}(v, w) \rangle_{\rho} \neq \int \text{d}v\int\text{d}w\, p(v, w) \mathcal{O}(v, w)\,.
	\end{equation}
	Instead, only the visible states incorporate the degrees of freedom of the originally considered expectation value.
	
	We note that, mathematically, this is clear in the case of complex Langevin dynamics for a single complex field $\phi = \phi_x + \iu \phi_y$. The original integral over $\phi$ is one-dimensional rather than a two-dimensional surface integral over the complex plane. As a result, the sum in Eq.~\eq{eq:NumExpectationValueGeneral} returns the mean of the integral for different values of the hidden variables and thus an expectation value with respect to the original distribution $\rho(x)$.
	
	This computation of expectation values with auxiliary variables differs significantly from existing ones. We emphasise that $v = x$ does not hold and a marginalization over $w$ is absent. In the following, we discuss, besides complex Langevin dynamics, several algorithms that implement the above principles and satisfy all of the given constraints for this approach. Keeping all the constraints in mind, one can make use of general relations and methods for sampling from high-dimensional probability distributions, such as Markov chain Monte Carlo methods.
	
	\section{Substitution sampling}
	\label{sec:SubstituionSampling}
	
	In this chapter, we formulate in Sec.~\ref{sec:GeneralDefinition} the general constraints a sampling algorithm needs to fulfil which serve to compute expectation values of the kind defined in Eq.~\eq{eq:NumExpectationValueGeneral}. We will refer to this kind of sampling algorithm as \textit{Substitution Sampling} to reflect that it is built on the key insights in Sec.~\ref{sec:KeyFindings}. Additionally, we identify complex Langevin dynamics as such an algorithm and provide a guide for constructing substitution sampling algorithms in Sec.~\ref{sec:Construction}.
	
	\subsection{General definition}
	\label{sec:GeneralDefinition}
	
\begin{figure}
	\centering \includegraphics[width=\linewidth]{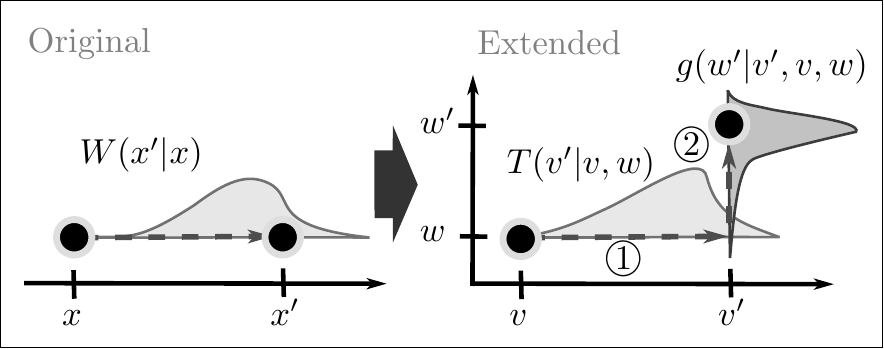}
	\caption{Schematic illustration of the transition probabilities (left) in the original and (right) the extended representation space. The visible variables $v$ are updated according to the transition probability $T(v'|v, w)$. The new hidden states $w'$ are obtained in a second step, involving the transition probability $g(w'|v', v, w)$.} \label{fig:DetailedBalance}
	\end{figure}

	The proposed substitution sampling algorithm generates dynamics in the set of variables $(v, w)$ as a Markov process with transition probabilities $W(v, w\to v', w')$. It distinguishes between an update step that only affects the visible variables and one that only changes the hidden variables. This is implemented by splitting the transition probability into conditional probabilities $T$ and $g$ for visible and hidden states, respectively. The splitting can be done in two ways, with an update first of the visible variables, followed by a conditional update of the hidden ones,
	\begin{equation}
	W(v, w\to v', w') =
	\,g(w'|v', v, w) T(v'|v, w)\,,
	\end{equation}
	or vice versa,
	\begin{equation}
	W(v, w\to v', w') = T(v'|v, w', w) g(w'|, v, w)\,.
	\end{equation}
	In the following, we will only use the first splitting, although both variants are possible. The differences between an update step in the original state space and one in the higher-dimensional state space are schematically shown in Fig.~\ref{fig:DetailedBalance}. One update step consists of a sequential update of the visible and the hidden states.
	
	A substitution sampling algorithm needs to satisfy, in the large-time limit, the following constraints:
	\begin{enumerate}
		\item \label{item:1} Satisfaction of the following detailed-balance equation for a fixed hidden state $w$:
		\begin{align}
		\label{eq:DetailedBalanceSubstition}
		p(v&, w) g(w'|v', v, w) T(v' | v, w) \nonumber\\[1ex]&= p(v', w) g(w'|v, v', w) T(v | v', w)\,.
		\end{align}
		\item \label{item:2} The hidden states $w$ are updated with an infinitesimal step size.
		\item \label{item:3} The mean values~\eq{eq:SamplingGeneral} are invariant under shifts of the boundaries $a$ and $b$ by any of the sampled hidden state variables $w_i$. This is satisfied, for example, if $p(v, w)$ converges sufficiently fast to zero near the integral boundaries.
		\item \label{item:4} The distribution $p(v, w)$ and the transition probabilities $T$ and $g$ need to satisfy the constraint, cf. Eq.~\eq{eq:EquilibriumConstraint},
		\begin{align}
		\label{eq:EquilibriumConstraintIntegral2}
		p(v',& w',\tau)\stackrel{!}{=} \int\text{d}v\int\text{d}w \nonumber\\[1ex] &\, \times p(v, w,\tau)\, g(w'|v', v, w) T(v'|v, w)\,.
		\end{align}
	\end{enumerate}
	
	The first two constraints ensure that the hidden states do not introduce any stochastic behaviour with respect to the distribution $p(v, w)$ and that only the visible states incorporate the degrees of freedom of the originally considered expectation value. In numerical simulations, it can also be sufficient if the stochastic behaviour in the visible direction dominates the one in the hidden direction. This is, for example, the case for complex Langevin dynamics with imaginary noise~\cite{Aarts2010, Aarts2013} and for the algorithms discussed in Sec.~\ref{sec:UniformComplexLangevin} and Chapter~\ref{sec:SHMCS}.
	
	The last constraint enforces that the substitution sampling algorithm, at long times, formally, samples from the equilibrium distribution $p(v, w)$. As pointed out above, it is feasible to make use of the condition~\eq{eq:EquilibriumConstraintIntegral2} since relations of Monte Carlo sampling algorithms in higher dimensions can be used for a computation of observables according to Eq.~\eq{eq:NumExpectationValueGeneral} as long as the hidden variables introduce no stochastic contribution to the computed expectation value. Because of this, the actually observed distribution differs from $p(v, w)$. Instead, numerical observables coincide with expectation values with respect to the underlying distribution $\rho(x)$:
	\begin{equation}
	\langle \mathcal{O}(x) \rangle_{\rho} = \langle \mathcal{O}(v + w) \rangle_{p}\,.
	\end{equation}
	In the case of a complex probability measure $p(v, w)$ the transition probabilities $T$ and $g$ need to be real-valued and positive to allow an actual sampling. In the next section, we show how this is implemented for complex Langevin dynamics.
	
	\subsection{Complex Langevin as a substitution sampling algorithm}
	\label{sec:ComplexLangevinAsASubstitutionSamplingAlgorithm}
	
	We show in this section that complex Langevin dynamics can be attributed to the class of substitution sampling algorithms. But before that, we want to point out that the complex Langevin equations can also be systematically derived by imposing the respective constraints for complex actions, as worked out explicitly in App.~\ref{sec:LangevinSamplingByCompensation}. The algorithms discussed in Chapter~\ref{sec:FurtherLangevinLikeAlgorithms} are derived in the same way.
	
	The approach allows deriving transition probabilities for complex Langevin dynamics. We use these transition probabilities in the following to prove a satisfaction of constraints no.~\ref{item:1} to no.~\ref{item:4}.

	\subsubsection*{Transition probabilities}
	
	In concordance with the discussion in Chapter~\ref{sec:MainContributions} and in the previous section, our goal is to show that complex Langevin dynamics, formally, samples from the complex distribution
	\begin{equation}
	\label{eq:Rho}
	\rho(\phi_x + \iu \phi_y) \propto \exp(-S(\phi_x + \iu \phi_y))\,
	\end{equation}
	while satisfying the constraints no.~\ref{item:1} to no.~\ref{item:4} as required for a substitution sampling algorithm. The constraints demand that the stochastic contribution in the $\phi_y$ direction vanishes in a certain limit. In the following, we will specify this limit for the case of complex Langevin dynamics for which it is reached with an evolution in the continuous Langevin time $\tau$.
	
	We thereby assume that constraint no.~\ref{item:3}, namely an invariance under simultaneous shifts of the integration boundaries, is satisfied by the considered observables, which holds independently of the transition probabilities.
	
	Driven by the motivation to view complex Langevin dynamics from the perspective of a Markov chain Monte Carlo algorithm, we start with a discretization of the Langevin time in Eq.~\eq{eq:ComplexLangevin},
	\begin{align}
	\label{eq:IntroComplexLangevin} 
	\phi_x' &= \phi_x - \epsilon\,\text{Re}\left[\frac{\delta S(\phi)}{\delta \phi}\bigg|_{\phi_x
		+ \iu \phi_y}\right] + \sqrt{2\epsilon} \eta\,,\nonumber\\[1ex]
	\phi_y' &= \phi_y - \epsilon\,
	\text{Im}\left[\frac{\delta S(\phi)}{\delta \phi}\bigg|_{\phi_x
		+ \iu \phi_y}\right]\,,
	\end{align}
	where $\epsilon=\Delta \tau$ is the time step in which $\phi_x$ and $\phi_y$ evolve to $\phi_x'$ and $\phi_y'$. This formulation allows a numerical implementation of the evolution. The continuous limit in the Langevin time ($\epsilon\to 0$) is evaluated by extrapolating the results of repeated simulations for different values of $\epsilon$.

	The update rule for the real part in Eq.~\eq{eq:IntroComplexLangevin} can be obtained by means of an expansion of the real part of the action difference $\Delta S_{\text{Re}}(\phi', \phi) = S_{\text{Re}}(\phi_x'+ \iu\phi_y) - S_{\text{Re}}(\phi_x + \iu\phi_y)$ in the transition probability,
	\begin{align}
	\label{eq:ResultingTransitionComplexLangevinS}
	T(\phi_x'&|\phi_x, \phi_y)\nonumber\\[1ex]&\propto \varphi\left(\frac{\phi_x'-\phi_x}{\sqrt{2\epsilon}}\right)\exp\left(-\frac{\Delta S_{\text{Re}}(\phi', \phi)}{2}\right)\,.
	\end{align}
	This expression for the transition probability is derived in App.~\ref{sec:LangevinSamplingByCompensation}, cf. Eq.~\eq{eq:ResultingTransitionComplexLangevin}. Here, $\varphi$ denotes the Gaussian distribution
	\begin{equation}
	\label{eq:GaussianDistribution}
	\varphi(x) = \frac{1}{\sqrt{2\pi}}\exp\left(-\frac{x^2}{2}\right)\,.
	\end{equation}
	Hence, the transition probability $T$ is given by the product of a proposal distribution $\varphi$ for the new field value $\phi_x'$ and an acceptance probability that depends on the action difference $\Delta S_{\text{Re}}(\phi', \phi)$.
	
	Analogously, the update equation for the imaginary part of complex Langevin dynamics, see Eq.~\eq{eq:IntroComplexLangevin}, involves the imaginary part of the action difference $\Delta S_{\text{Im}}(\phi', \phi) = S_{\text{Im}}(\phi_x'+ \iu\phi_y) - S_{\text{Im}}(\phi_x + \iu\phi_y)$, cf. Eq.~\eq{eq:UpdatePhiY},
	\begin{equation}
	\label{eq:UpdatePhiYS}\phi_y' = \phi_y - \epsilon
	\frac{\Delta S_{\text{Im}}(\phi', \phi)}{\phi_x' - \phi_x} \,.
	\end{equation}
	Since it does not contain any noise term, the respective conditional transition probability is a delta-distribution,
	\begin{equation}
	\label{eq:UpdatePhiYg}
	g(\phi_y'|\phi_x', \phi_x, \phi_y) = \delta\left(\phi_y' - \phi_y + \epsilon
	\frac{\Delta S_{\text{Im}}(\phi', \phi)}{\phi_x' - \phi_x}\right)\,.
	\end{equation}
	The transition probability $g$ defines the update rule for $\phi_y'$, where we use that $x=\phi$, $v=\phi_x$ and $w=\iu\phi_y$. An expansion of the action difference to first order yields the update equation of the imaginary part of complex Langevin dynamics. In the limit $\epsilon\to 0$, constraint no.~\ref{item:2}, demanding an infinitesimal step size into the $\phi_y$ direction, is thus obeyed.
	
	Note that, in the derivation of both update rules, the action difference involves a change in $\phi_x$ only. This is in accordance with the condition that only the visible variables represent the degrees of freedom of the initially considered expectation value over $x$.
	
	The derivation of the discrete update equations~\eq{eq:IntroComplexLangevin} is performed explicitly in App.~\ref{sec:ComplexLangevinDynamicsCLELike}, starting from the transition probabilities~\eq{eq:ResultingTransitionComplexLangevinS} and~\eq{eq:UpdatePhiYg}.
	
	\subsubsection*{Langevin symmetry}
	
\begin{figure}
	\centering \includegraphics[width=\linewidth]{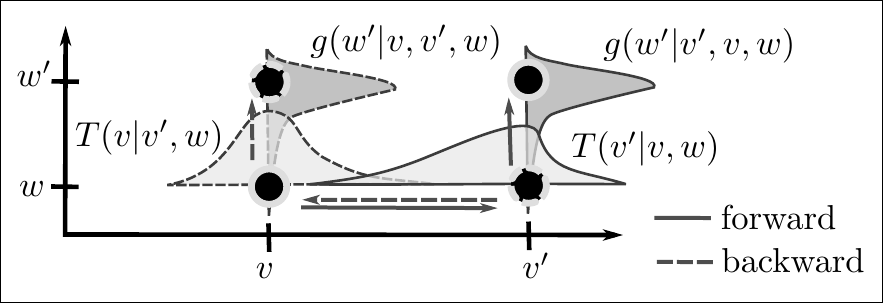}
	\caption{Illustration of the Langevin symmetry defined in Eqs.~\eq{eq:LangevinSymmetryCLE} and~\eq{eq:SymmetryLangevin}. The transition probabilities for the visible variables in the forward and the backward directions of the adapted detailed-balance equation~\eq{eq:DetailedBalanceT} are different. In contrast, the transition probability for the hidden state $w$ is invariant under an exchange of $v'$ and $v$.}  \label{fig:Symmetry}
	\end{figure}

	We point out that the transition probability~\eq{eq:UpdatePhiYg} for the imaginary part is invariant under an exchange of $\phi_x'$ and $\phi_x$,
	\begin{equation}
	\label{eq:LangevinSymmetryCLE}
	g(\phi_y'|\phi_x', \phi_x, \phi_y) = g(\phi_y'|\phi_x, \phi_x', \phi_y)\,.
	\end{equation}
	We will refer to this symmetry as \textit{Langevin symmetry}, which will be a key ingredient for the construction of substitution sampling algorithms. See Fig.~\ref{fig:Symmetry} for an illustration of the symmetry.

	If the Langevin symmetry holds, constraint no.~\ref{item:1} reduces to
	\begin{equation}
	\label{eq:DetailedBalanceLangevin}
	p(\phi_x, \phi_y) T(\phi_x' |\phi_x, \phi_y) \stackrel{!}{=} p(\phi_x', \phi_y)  T(\phi_x | \phi_x', \phi_y)\,,
	\end{equation}
	where $p(\phi_x, \phi_y) = \rho(\phi_x + \iu \phi_y)$, cf. Eq.~\eq{eq:Rho}.
	
	In fact, the transition probability $T$, defined in Eq.~\eq{eq:ResultingTransitionComplexLangevinS}, violates this modified detailed-balance equation, since:
	\begin{align}
	\label{eq:DetailedBalanceLangevinAdapted}
	p(&\phi_x, \phi_y) T(\phi_x' |\phi_x, \phi_y) \nonumber\\[1ex] &  = p(\phi_x', \phi_y)  T(\phi_x | \phi_x', \phi_y)\exp(-\iu\Delta S_\text{Im}(\phi, \phi'))\,.
	\end{align}
	However, Eq.~\eq{eq:DetailedBalanceLangevinAdapted} is satisfied if the step size into the $\phi_x$ direction is also chosen to be infinitesimal. This is ensured in the limit $\epsilon\to 0$ since the proposal distribution converges to a delta-distribution around $\phi_x$,
	\begin{equation}
	\lim\limits_{\epsilon\to 0}\frac{1}{\sqrt{2\epsilon}}\varphi\left(\frac{\phi_x'-\phi_x}{\sqrt{2\epsilon}}\right) = \delta (\phi_x' - \phi_x)\,.
	\end{equation}
	The infinitesimal step size in the $\phi_x$ direction justifies the previously performed expansion in the action difference and ensures constraint no.~\ref{item:4} to be fulfilled:
	\begin{align}
	\label{eq:ConstraintFourCLE}
	p(\phi_x',& \phi_y',\tau)\stackrel{!}{=} \int\text{d}\phi_x\int\text{d}\phi_y \nonumber\\[1ex] &\, \times p(\phi_x, \phi_y,\tau)\, g(\phi_y'|\phi_x', \phi_x, \phi_y) T(\phi_x'|\phi_x, \phi_y)\,.
	\end{align}
	This is proven as follows. We start by inserting relation~\eq{eq:DetailedBalanceLangevinAdapted} into Eq.~\eq{eq:ConstraintFourCLE}. We then make use of the symmetry~\eq{eq:LangevinSymmetryCLE} and finally expand the action difference $\Delta S_{\text{Im}}(\phi, \phi')$ to first order around $\phi_x'$, which gives
	\begin{align}
	\label{eq:LangevinConstraint}
	p(&\phi_x', \phi_y',\tau)\stackrel{!}{=} \int\text{d}\phi_x\int\text{d}\phi_y\,p(\phi_x', \phi_y,\tau)\, g(\phi_y'|\phi_x', \phi_y) \nonumber\\[1ex] & \times\,T(\phi_x|\phi_x', \phi_y)\,\exp\left(-\iu(\phi_x - \phi_x') \frac{\delta S_\text{Im}(\phi_x' + \iu\phi_y)}{\partial \phi_x'}\right)\,.
	\end{align}
	Here, $g(\phi_y'|\phi_x', \phi_y) \equiv g(\phi_y'|\phi_x, \phi_x', \phi_y)$, i.e., the $\phi_x$-dependence of the transition probability $g$ can be dropped due to the expansion of $\Delta S_\text{Im}(\phi, \phi')$ to first order around $\phi_x'$. The expansion is justified in the limit $\epsilon\to 0$, where also $\phi_x$ changes by infinitesimal amounts only. In this limit, it is possible to absorb the exponential function in Eq.~\eq{eq:LangevinConstraint} into the transition probability $T(\phi_x|\phi_x', \phi_y)$. See App.~\ref{sec:AbsorbingTheImaginaryContribution} for further details.
	
	We can now integrate over $\phi_x$ since the transition probability on the right-hand side is the only distribution depending on $\phi_x$ and, using its normalization, we are left with
	\begin{equation}
	\label{eq:FinalConstraint}
	p(\phi_x',\phi_y',\tau)\stackrel{!}{=} \int\text{d}\phi_y\, p(\phi_x', \phi_y,\tau)\, g(\phi_y'|\phi_x', \phi_y)\,.
	\end{equation}
	As a last step, we take the limit $\epsilon\to 0$. In this limit, by the definition of the conditional transition probability $g$, Eq.~\eq{eq:FinalConstraint} is indeed satisfied by $p(\phi_x, \phi_y) = \lim\limits_{\tau \to \infty}p(\phi_x, \phi_y, \tau)$. This completes the proof.
	
	We conclude that the step sizes in configuration space need to be infinitesimal in both the $\phi_x$ and the $\phi_y$ directions in order to fulfil the constraints no.~\ref{item:1} to no.~\ref{item:4} required for a substitution sampling algorithm. Hence, for $\epsilon\to0$, the transition probabilities~\eq{eq:ResultingTransitionComplexLangevinS} and~\eq{eq:UpdatePhiYg} are equivalent to the discretized update rules~\eq{eq:IntroComplexLangevin} and thus to complex Langevin dynamics.
	
	\subsection{Constructing substitution sampling algorithms}
	\label{sec:Construction}

	In the following, we generalize the key concepts of complex Langevin dynamics discussed in the previous section to the case of general visible and hidden variables and thus also to finite step sizes in the visible direction. This generalization provides a possible approach to constructing transition probabilities that satisfy all of the constraints a substitution sampling algorithm must fulfil.
	
	We start again by demanding that the transition probability for the hidden variables obeys the Langevin symmetry (see also Fig.~\ref{fig:Symmetry})
	\begin{equation}
	\label{eq:SymmetryLangevin}
	g(w' | v', v, w) = g(w' | v, v', w)\,.
	\end{equation}
	With this symmetry, the detailed-balance equation~\eq{eq:DetailedBalanceSubstition} can be written as
	\begin{align}
	\label{eq:DetailedBalanceT}
	g&(w' | v', v, w)\nonumber\\[1ex]& \times \left[p(v, w) T(v' | v, w) - p(v', w)  T(v | v', w)\right] \stackrel{!}{=} 0\,.
	\end{align}
	For non-vanishing $g$, the term in square brackets, referred to as adapted detailed-balance equation, must vanish, which constrains the transition probabilities $T(v'|v,w)$.
	
	The meaning of the adapted detailed-balance equation becomes clearer when one takes a closer look at the equation: It can be viewed as a detailed-balance equation of a Markov chain that allows changes in the visible state variables $v$ only, whereby $w$ is fixed. The process is unaware of any dependence on the additional auxiliary variables $w$. Nevertheless, $w$ will be updated based on $g(w'|v', v, w)$. This entails a transformation of the environment for the Markov chain in $v$ after each update step since the action depends on $w$.
	
	The above interpretation mirrors the important concept of the substitution sampling algorithm for computing observables by means of an integration over the visible variables only. In contrast, the hidden variables give rise to a continuous set of different substitutions of the dynamical variables in the originally considered integral and carry no stochastic behaviour, cf. Eq.~\eq{eq:SamplingGeneral}.
	
	It remains to derive transition probabilities $g$ for the hidden states that are in concordance with the constraints no.~\ref{item:2} to no.~\ref{item:4}.
	
	In the following, we point out possible implications that result from constraint no.~\ref{item:4}, Eq.~\eq{eq:EquilibriumConstraintIntegral2}. We insert the detailed-balance equation~\eq{eq:DetailedBalanceSubstition} into the right-hand side of Eq.~\eq{eq:EquilibriumConstraintIntegral2},
	\begin{align}
	\label{eq:ConstraintFourGen}
	p(&v', w', \tau) = \int\text{d}v\int\text{d}w\, g(w'|v', v, w) T(v'|v, w) p(v, w,\tau) \nonumber\\[1ex] &= \int\text{d}v\int\text{d}w\,g(w'|v, v', w) T(v|v', w) p(v', w,\tau)\,.
	\end{align}
	Inspired by the first-order expansion~\eq{eq:LangevinConstraint} in the case of complex Langevin dynamics, we here demand that $g$ does not depend on $v$,
	\begin{equation}
	g(w'|v, v', w) \equiv g(w'|v', w)\,.
	\end{equation}
	As for complex Langevin dynamics, this allows performing the integration over $v$ in Eq.~\eq{eq:ConstraintFourGen}, resulting in
	\begin{equation}
	\label{eq:HiddenEquilibriumConstraint}
	p(v', w',\tau) \stackrel{!}{=}\int\text{d}w\,g(w'|v', w) p(v', w,\tau)\,.
	\end{equation}
	Next, we make use of constraint no.~\ref{item:2}, which suggests that $g$ is of the form
	\begin{equation}
	\label{eq:HiddenTransitionConstraint}
	g(w'|v', w) = \delta\left(w' - h(v', w; \epsilon)\right)\,,
	\end{equation}
	where $\delta(\cdot)$ represents the delta-distribution and the function $h(v', w; \epsilon)$ has the property that
	\begin{equation}
	\lim\limits_{\epsilon\to 0} h(v', w; \epsilon) = w\,.
	\end{equation}
	Here, the parameter $\epsilon$ parametrizes the step size in the update process of the hidden states. With the above assumptions on $g$, one can take the limit $\epsilon\to 0$ and integrate over $w$, which confirms constraint~\eq{eq:HiddenEquilibriumConstraint} to hold and therefore constraint no.~\ref{item:4}, cf. Eq.~\eq{eq:EquilibriumConstraintIntegral2}.
	
	Note that for a transition from $(v, w) \to (v', w')$, one needs to replace $v'$ by $v$ in Eq.~\eq{eq:HiddenTransitionConstraint},
	\begin{equation}
	\label{eq:HiddenTransitionConstraint2}
	g(w'|v, w) = \delta\left(w' - h(v, w; \epsilon)\right)\,.
	\end{equation}
	Complex Langevin dynamics deviates from this construction in the sense that the adapted detailed-balance equation~\eq{eq:DetailedBalanceT} is only warranted when step sizes into the visible direction are infinitesimal, too.
	
	In the next chapter, we introduce examples of algorithms that are constructed based on the same principles as complex Langevin dynamics. Chapter~\ref{sec:SHMCS} then provides an example of an algorithm that satisfies the constraints of a substitution sampling algorithm in a different way.

	\section{Complex Langevin-type algorithms}
	\label{sec:FurtherLangevinLikeAlgorithms}
	
	The analysis of complex Langevin dynamics and the above guide for constructing substitution sampling algorithms can be combined to define a systematic approach to deriving transition probabilities $T$ and $g$. The resulting algorithms differ in their proposal distributions and satisfy the constraints of substitution sampling in the same manner as complex Langevin dynamics.
	
	This systematic approach is described in detail in App.~\ref{sec:LangevinSamplingByCompensation}. It is inspired by an alternative derivation of complex Langevin dynamics which recovers known results from a different point of view. The core concepts of the derivation are: an extension of the transition probabilities of Langevin dynamics to a higher-dimensional state space and a compensation of certain (here imaginary) contributions in the action by terms that emerge in the transition to the extended state space. The approach is built on the requirement that the Langevin symmetry as well as constraints no.~\ref{item:1} to no.~\ref{item:4} stated in Sec.~\ref{sec:GeneralDefinition} are obeyed. It then leads to the transition probabilities of, e.g., complex Langevin dynamics, cf. Eqs.~\eq{eq:ResultingTransitionComplexLangevinS} and~\eq{eq:UpdatePhiYg}.
	
	The cancellation of imaginary contributions of the action is a crucial step in this derivation, cf. Eqs.~\eq{eq:ContributionTerm} and~\eq{eq:UpdateY}. In the case of complex Langevin dynamics, the update of $\phi_y'$ of the imaginary field $\phi_y$ is used for this. This compensation leads to well-defined, real-valued transition probabilities and, therefore, allows an actual sampling of problems with a sign problem.
	
	In the following, we present several algorithms resulting from the systematic approach. Detailed derivations of these algorithms are given in App.~\ref{sec:CLELikeAlgorithms}.
	
	\subsection{Second-order complex Langevin}
	\label{sec:SecondOrderComplexLangevin}
	
	\textit{Second-order complex Langevin dynamics} results as a refinement of complex Langevin dynamics. For this, also the
	second-order term of the Taylor expansion, cf. Eq.~\eq{eq:Taylor}, of the action
	difference around $\phi_x$ is taken into account. The resulting update rule for the real part
	of the field is %
	\begin{equation}
	\label{eq:CLESecondOrderReal}
	\phi_x' = \phi_x - \left(\epsilon \frac{\delta
		S_{\text{Re}}}{\delta \phi_x} + \sqrt{2\epsilon} \eta\right)\bigg/\left(1 +
	\frac{\epsilon}{2} \frac{\delta^2 S_{\text{Re}}}{\delta \phi_x^2}\right)\,,
	\end{equation}
	and, for the imaginary part,
	\begin{equation}
	\label{eq:CLESecondOrderImag}
	\phi_y' = \phi_y - \epsilon \frac{\delta
		S_{\text{Im}}}{\delta \phi_x} - \frac{\epsilon}{2}\left(\phi_x' -
	\phi_x\right) \frac{\delta^2 S_{\text{Im}}}{\delta \phi_x^2}\,,
	\end{equation}
	where we defined $S_{\text{Re}}:=S_{\text{Re}}(\phi_x + \iu\phi_y)$
	and $S_{\text{Im}}:=S_{\text{Im}}(\phi_x + \iu\phi_y)$. As before, the
	update rule samples from the desired
	equilibrium distribution in the limit of $\epsilon \to 0$  since detailed balance is satisfied only in this
	limit. Details on the derivation can be found in
	App.~\ref{sec:AppendixSecondOrderComplexLangevin}. A numerical comparison to
	complex Langevin dynamics will be presented in Chapter~\ref{sec:NumericalResults}.
	
	\subsection{Complex hat function algorithm}
	\label{sec:ComplexHatFunctionAlgorithm}
	
	Complex Langevin dynamics uses a Gaussian distribution $\varphi$, cf. Eq.~\eq{eq:GaussianDistribution}, in proposing states $\phi'$. We demonstrate, in this section, that the systematic derivation of Langevin-type sampling algorithms does also work for other types of proposal distributions. In particular, we consider the triangular hat function,
	\begin{equation}
	\eta_\epsilon(\phi' - \phi) = \frac{1}{\epsilon} \begin{cases}
	1-\frac{\phi'-\phi}{\epsilon}\quad&\textnormal{for}\;0\,\leq\, \phi'-\phi \,<\, \epsilon\,,\\
	1+\frac{\phi'-\phi}{\epsilon}\quad&\textnormal{for}\;-\epsilon\,<\, \phi'-\phi \,<\,0\,,\\
	0\quad&\quad\text{otherwise.}
	\end{cases}
	\end{equation}
	as a proposal distribution. The limit $\epsilon\to 0$ facilitates the implementation of an infinitesimal step size in configuration space. This is a necessary condition to satisfy the constraints of the substitution algorithm, as worked out in App.~\ref{sec:LangevinSamplingByCompensation}.
	
	We assume again a complex action, as defined for the polynomial model in Eq.~\eq{eq:PolynomialModel}, and define an update scheme that allows sampling despite a sign problem. The respective update rules for $\phi_x$ and $\phi_y$ are derived with the help of the systematic derivation in App.~\ref{sec:AppendixComplexHatFunctionAlgorithm}.
	
	The update rule for the imaginary field $\phi_y$ is given by:
	\begin{equation}
	\label{eq:HatImag}
	\phi_y' = \phi_y+ \left[\frac{\epsilon}{s} - \left(\phi_x' - \phi_x\right)\right]\tan\left(-\frac{\Delta S_\text{Im}(\phi', \phi)}{2}\right)\,,
	\end{equation}
	where
	\begin{equation}
	s = \text{sign}(\phi_x' - \phi_x)\,.
	\end{equation}
	The update rule is invariant under an exchange of $\phi_x'$ and $\phi_x$ and thus possesses the Langevin symmetry, Eq.~\eq{eq:SymmetryLangevin}.
	
	The update rule compensates the contributions from the imaginary part of the action as it is also the case for complex Langevin dynamics. This compensation leads to a real-valued transition probability for the real part of the field $\phi_x$, namely,
	\begin{align}
	\label{eq:HatReal}
	T(&\phi_x'|\phi_x, \phi_y) =
	\frac{1}{\epsilon N(\phi)}\exp\left(-\frac{\Delta S_\text{Re}(\phi', \phi)}{2}\right)\nonumber\\[1ex]&\times \left(1 - s \frac{\phi_x' - \phi_x}{\epsilon}\right) \cos^{-1}\left(-\frac{\Delta S_\text{Im}(\phi', \phi)}{2}\right)\,.
	\end{align}
	In contrast to complex Langevin dynamics, it is not trivial to translate this transition probability in an update rule for $\phi_x$. Instead, we sample a new state $\phi_x'$ implicitly by numerically solving the transformation of the transition probability to a uniform distribution,
	\begin{equation}
	\label{eq:TransformationTransitionProbabilityComplexHat}
	\int_{-\infty}^{\phi_x'} \text{d}\tilde{\phi}_x \, T(\tilde{\phi}_x|\phi_x, \phi_y)
	\stackrel{!}{=} \int_{0}^r \text{d}\tilde{r} = r\,.
	\end{equation}
	In practice, one samples $r$ from the uniform distribution and numerically solves the expression on the left-hand side for $\phi_x'$, so that the equality is satisfied for the sampled $r$. It is important that the transition probability $T$ represents a probability distribution. In the limit of $\epsilon\to0$ this is indeed the case.
	
	\subsection{Uniform complex Langevin}
	\label{sec:UniformComplexLangevin}
	
	A substitution sampling algorithm can also be formulated for a uniform proposal distribution. We achieve this by defining the proposal distribution by means of an integrated delta-distribution
	\begin{align}
	q(\phi&\to\phi') = \int_{-l}^{l} \frac{\text{d}r}{2l}\,\delta\left(\phi' - (\phi + r)\right)\,\nonumber\\[1ex]&= \frac{1}{2l}\left[\Theta\left(\phi' - \phi + l\right) - \Theta\left(\phi' - \phi - l)\right)\right]\,,
	\end{align}
	and implement it by sampling $r$ uniformly from the interval~$\left[-l, l\right]$.
	
	The resulting update rules are
	\begin{align}
	\label{eq:UniformComplexLangevinReal}
	T&(\phi_x'|\phi_x, \phi_y)\propto \int_{-l}^{l} \frac{\text{d}r}{2l}\,\delta\left(\phi_x' - (\phi_x + r)\right)\nonumber\\[1ex]&\times\exp\left(-\frac{\Delta S_{\text{Re}}(\phi', \phi)}{2}\right)	
	 \cos^{-1}\left(-\frac{\Delta S_\text{Im}(\phi', \phi)}{2}\right)
	\end{align}
	for the real part $\phi_x$ and
	\begin{equation}
	\label{eq:UniformComplexLangevinImag}
	\phi_y' = \phi_y + \left(\tilde{\phi}_x' - (\phi_x + r)\right)\tan\left(-\frac{\Delta S_\text{Im}(\phi', \phi)}{2}\right)
	\end{equation}
	for the imaginary part $\phi_y$ of the field. Sampling a state $\phi_x'$ works in the same manner as for the complex hat function algorithm by a transformation of the transition probability, cf. Eq.~\eq{eq:TransformationTransitionProbabilityComplexHat}. In contrast to the other approaches, two proposal states, $\phi_x'$ and $\tilde{\phi}_x'$ are sampled. This entails a finite step size for $\phi_y$. The algorithm satisfies all constraint of a substitution algorithm in the limit of $l\to 0$, see App.~\ref{sec:LangevinSamplingByCompensation} for details.
	
	In principle, any other proposal distribution can be used as long as constraint~\eq{eq:HiddenTransitionConstraint} for the imaginary update rule is satisfied and potentially introduced noise in the imaginary direction is dominated, in numerical simulations, by the noise in the real direction in the limit of infinitesimally small step sizes.
	
	\subsection{Metropolis-like sampling}
	
	In principle, it is also possible to define a Metropolis accept/reject step based on the adapted detailed-balance equation~\eq{eq:DetailedBalanceLangevin}. The acceptance probability is approximated by
	\begin{align}
	\label{eq:MetropolisLikeSampling}
	A(&\phi_x'|\phi_x, \phi_y) \nonumber\\[1ex] & = \min\left[1, \exp\left(-(S_\text{Re}(\phi_x' + \iu\phi_y) - S_\text{Re}(\phi_x + \iu\phi_y)\right)\right]\,.
	\end{align}
	where a respective transition probability $T$ is defined as the product of a symmetric proposal distribution for $\phi_x'$ and the acceptance probability according to:
	\begin{equation}
	T(\phi_x'|\phi_x, \phi_y) = q(\phi_x'|\phi_x)A(\phi_x'|\phi_x, \phi_y)\,.
	\end{equation}
	The adapted detailed-balance equation is violated by this definition in the same way as for complex Langevin dynamics and the other complex Langevin-type algorithms in this chapter, cf. Eq.~\eq{eq:DetailedBalanceLangevinAdapted}. Accordingly, the sampling algorithm works also only in the limit of infinitesimally small step sizes in $\phi_x$. The imaginary part $\phi_y$ is updated based on the associated update equation $g$ of the used proposal distribution, independent of an acceptance or a rejection of the proposed state.
	
	\section{Substitution Hamiltonian Monte Carlo sampling in auxiliary dimensions}
	\label{sec:SHMCS}
	
	We present, in this section, as a proof of concept, an alternative algorithm satisfying constraints no.~\ref{item:1} to no.~\ref{item:4} of a substitution sampling algorithm defined in Sec.~\ref{sec:GeneralDefinition}. We call the algorithm \textit{Substitution Hamiltonian Monte Carlo Sampling} (SHMCS). The algorithm is not derived within the systematic approach introduced in App.~\ref{sec:LangevinSamplingByCompensation}. Instead, it makes use of the basic idea of the Hamiltonian Monte Carlo algorithm to introduce an additional momentum as an auxiliary dimension. The SHMCS algorithm only works for real actions and cannot be applied to problems with a sign problem. However, it serves as a good example and provides numerical evidence in support of the general framework introduced in this work.
	
	We consider a (real) probability distribution $\rho(x)$ and want to design an algorithm for computing observables according to Eq.~\eq{eq:NumExpectationValueGeneral},
	\begin{equation}
	\label{eq:ExpectationValueSubstition2}
	\langle \mathcal{O}(x) \rangle = \langle \mathcal{O}(v + w) \rangle = \frac{1}{MN}\sum_{i=1}^{MN} \mathcal{O}(v_{i} + w_{i})\,.
	\end{equation}
	In contrast to complex Langevin dynamics, the hidden variables $w$ are taken to be real and introduced by the substitution
	\begin{equation}
	\label{eq:SubstitutionSHMCS}
	x = x(v) = v + w\,,\quad \text{d}x = \text{d}v\,.
	\end{equation}
	We assume that the probability distribution $p(v + w)$ satisfies the necessary constraints for a valid computation of expectation values by Eq.~\eq{eq:ExpectationValueSubstition2}.
	
	In addition, we introduce momenta $\pi$ as further hidden variables. Similar to the Hamiltonian Monte Carlo algorithm, the momenta are related to the visible variables $v$ through an energy function
	\begin{equation}
	H(v, w, \pi):= S(v + w) + \frac{\pi^2}{2m}\,.
	\end{equation}
	Next, we split the action into two contributions:
	\begin{equation}
	S(v + w) = S_1(v + w) + S_2(v + w)\,.
	\end{equation}
	This step is similar to the distinction between the real and the imaginary part of a complex action. For a real action, however, the splitting is arbitrary. 
	
	In contrast to the HMC algorithm, we demand
	\begin{equation}
	\tilde{H}(v, \pi):= S_2(v + w) + \frac{\pi^2}{2m}
	\end{equation}
	to stay constant during the Monte Carlo evolution which is implemented by updating $v$ and $\pi$ according to the differential equations
	\begin{equation}
	\label{eq:EvolutionSHMCS}
	\frac{\text{d} v}{\text{d} t} = \frac{\partial\tilde{H}(v, \pi)}{\partial \pi}\,,\quad \frac{\text{d} \pi}{\text{d} t} = -\frac{\partial\tilde{H}(v, \pi)}{\partial v}\,.
	\end{equation}
	which is assumed to be possible in a numerically exact manner. The remaining contribution $S_1(x)$ is taken into account through the acceptance term
	\begin{equation}
	A(v' | v, w) = \min\left[1, \exp\left(-\left(S_1(v' + w) - S_1(v + w)\right)\right)\right]\,.
	\end{equation}
	This approach satisfies, so far, the detailed-balance equation of a substitution sampling algorithm, cf. Eq.~\eq{eq:DetailedBalanceSubstition}. Therefore, we are free to choose an update rule for the transition probability of the hidden state $w$ as long as the constraints defined in Sec.~\ref{sec:GeneralDefinition} are satisfied. We define the transition probability $g$ as a Langevin process with finite step size,
	\begin{equation}
	\label{eq:SHMCSHiddenUpdate}
	g(w'|w) =  \varphi\left(\frac{w' - w}{\sqrt{2\epsilon}} + \sqrt{\frac{\epsilon}{2}}\,\theta w\right)\,,
	\end{equation}
	with the Gaussian distribution $\varphi$, cf. Eq.~\eq{eq:GaussianDistribution}. The transition probability translates in the limit $\epsilon\to 0$ in the Langevin evolution
	\begin{equation}
	\frac{\text{d} w}{\text{d} t} = -\theta w + \eta\,,
	\end{equation}
	with Gaussian noise $\eta$.
	
	Putting everything together, the SHMCS algorithm is defined based on the following transition probabilities
	\begin{align}
	\label{eq:SHMCS}
	T(v'|v, w, \pi)&\propto \delta\left(v' - R\,\Phi_{v}(v,w,
	\pi)\right) \nonumber\\[1ex]\times \min&\left[1, \exp\left(-(S_1(v'+ w) - S_1(v+
	w))\right)\right]\,,\nonumber\\[1ex] g(\pi'|v, w, \pi) &= \delta\left(\pi' - R\,\Phi_{\pi}(v, w, \pi)\right)\,,\nonumber\\[1ex]
	g(w'|w) &= \varphi\left(\frac{w' - w}{\sqrt{2\epsilon}} + \sqrt{\frac{\epsilon}{2}}\,\theta w\right)\,.
	\end{align}
	The functions $\Phi_{v}(v,w,
	\pi)$ and $\Phi_{\pi}(v,w,
	\pi)$ encode the end point of an evolution according to the differential equations~\eq{eq:EvolutionSHMCS} for a finite amount of time. The operator $R$ negates the momenta $\pi$ after the evolution. Similar to the HMC algorithm, this ensures reversibility.
	
	The transition probabilities satisfy all constraints of a substitution sampling algorithm. According to the fourth constraint, the SHMCS algorithm formally samples from the equilibrium distribution
	\begin{equation}
	p(v, w, \pi) \propto \exp\left(-H(v, w, \pi)\right)\,.
	\end{equation}
	As for complex Langevin dynamics, the actually observed steady-state distribution differs, due to the properties of the substitution sampling algorithm, from this distribution. The difference results from the required vanishing stochastic contribution in the direction of the hidden variables. In the case of the SHMCS algorithm, this requirement is violated by the Gaussian noise distribution in the transition probability $g$. However, these stochastic contributions are dominated in the limit of $\epsilon\to 0$ by finite correlations in the visible variables, resolving a violation of the requirement. In practice, this can be ensured by choosing $\theta$ sufficiently large.
	
	\section{Numerical results}
	\label{sec:NumericalResults}
	
	In the remainder of this work, we briefly examine the applicability of our approach and the algorithms derived with it by a numerical evaluation for the polynomial model defined in Eq.~\eq{eq:PolynomialModel},
	\begin{equation}
	\label{eq:PolynomialModel2}
	S(\phi) = \frac{1}{2}\left(\sigma_{\text{Re}} + \iu \sigma_{\text{Im}}\right)\phi^2 + \frac{\lambda}{4} \phi^4\,.
	\end{equation}
	Expectation values for benchmarking are analytically accessible for the chosen set of parameters, cf.~\cite{Aarts2013}.
	
\begin{figure*}
		\includegraphics{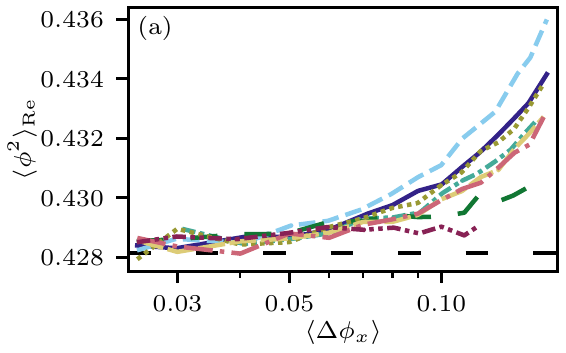}
		\hfill\includegraphics{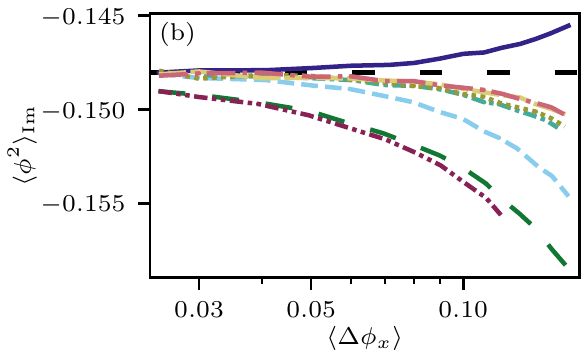}
		\hfill\includegraphics{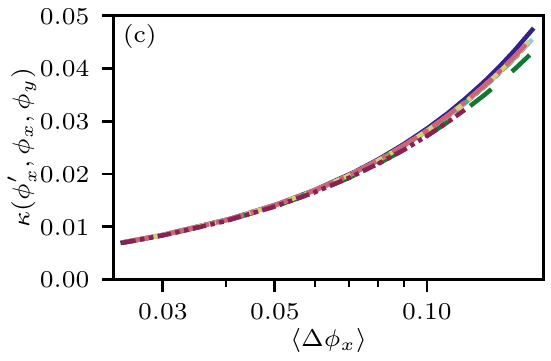}\\
		\includegraphics{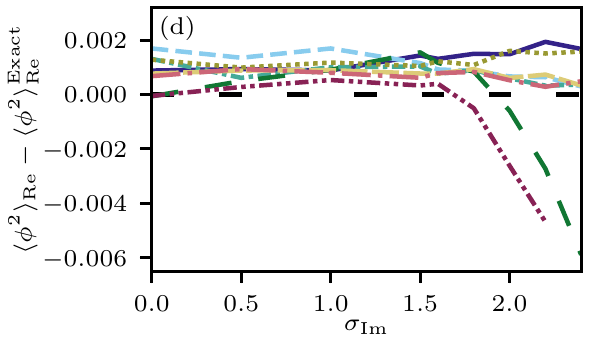}
		\hfill\includegraphics{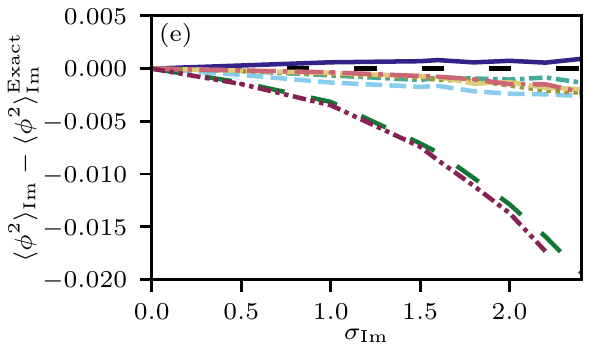}
		\hfill\vspace{-0.18cm}\includegraphics{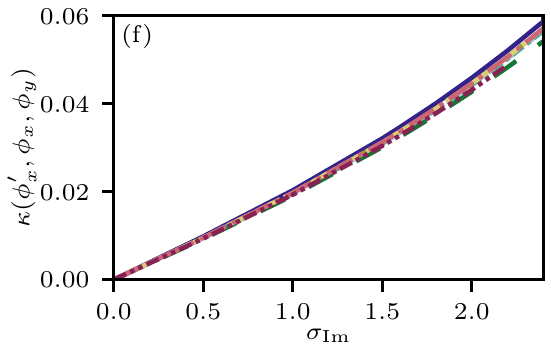}
		\includegraphics{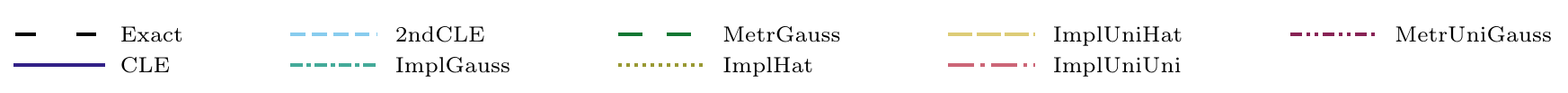}
		\caption{Comparison of numerical results for different complex Langevin-type sampling algorithms for the polynomial model~\eq{eq:PolynomialModel2}. Details about the algorithms are given in Table~\ref{tab:LangevinLikeAlgorithm}. (a)-(c) Results for $\lambda=\sigma_\text{Re}=\sigma_\text{Im}=1$ and a varying average step size $\langle \Delta \phi_x \rangle$ in the real direction of the representation space. (d)-(f) Results for $\lambda=\sigma_\text{Re}=1$ and a varying $\sigma_\text{Im}$. To get an appropriate comparison, the step sizes in the real direction were chosen to be equal for all algorithms. The plots (c) and (f) measure the violation of the adapted detailed-balance equation~\eq{eq:DetailedBalanceLangevin} based on the measure $\kappa(\phi_x', \phi_x, \phi_y)$ defined in Eq.~\eq{eq:AccuracyMeasure}. In concordance with Sec.~\ref{sec:ComplexLangevinAsASubstitutionSamplingAlgorithm}, the measure increases with the real step size and with the magnitude of the imaginary contribution, regulated by $\sigma_{\text{Im}}$.}
		\label{fig:ComparisonDynamics}
	\end{figure*}

\begin{figure} \includegraphics[width=0.97\linewidth]{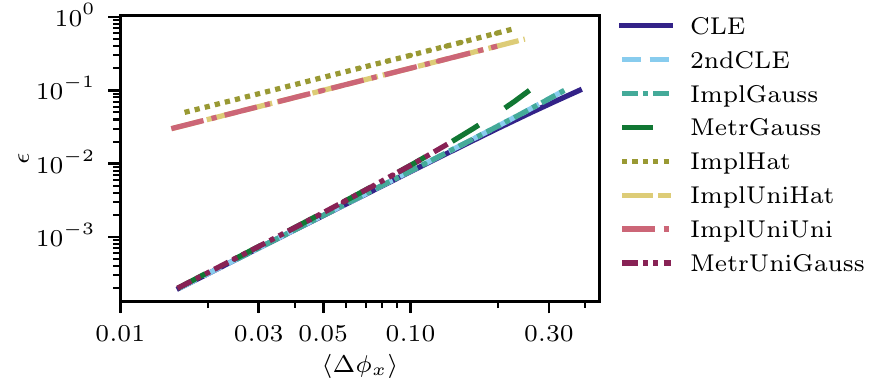}
		\caption{Relation between the parameter $\epsilon$ and the actually observed step size $\langle \Delta \phi_x \rangle$ in real direction for the different implemented complex Langevin-type algorithms.}
		\label{fig:RealStepSizes}	
	\end{figure}

	For the complex Langevin-type algorithms, we compare, in Fig.~\ref{fig:ComparisonDynamics}, the impact of finite step sizes on a possible extrapolation to the continuous limit and the performance for a fixed step size but a different severity of the sign problem, i.e., a more oscillating measure. The considered algorithms are defined in Table~\ref{tab:LangevinLikeAlgorithm}. The dependence of the measured average step size $\langle \phi_x \rangle$ in the real direction and on the chosen step size parameter $\epsilon$ is shown in Fig.~\ref{fig:RealStepSizes}.
	
	The results in Fig.~\ref{fig:ComparisonDynamics} show that none of the studied algorithms entails a significant difference to complex Langevin dynamics, with the exception of the Metropolis-like algorithms. The deviations can be traced back to the asymmetry between the accept and reject step for the real field variable and the independent update step of the imaginary field. The slightly worse convergence of second-order complex Langevin dynamics is likely related to an asymmetry in the adapted detailed-balance equation and the Langevin symmetry, cf. Eqs.~\eq{eq:LangevinSymmetryCLE} and~\eq{eq:DetailedBalanceLangevin}, introduced by the second order term of the Taylor expansion around $\phi_x$, cf. Eq.~\eq{eq:SecondOrderExpansion}.
	
\begin{figure*}
		\hspace{0.22cm}\includegraphics{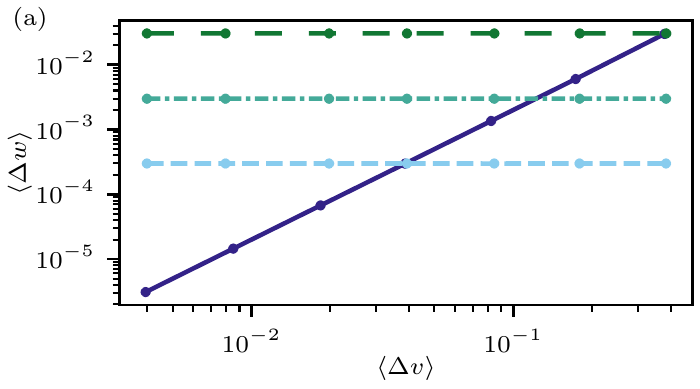}\hspace{0.0cm}\includegraphics{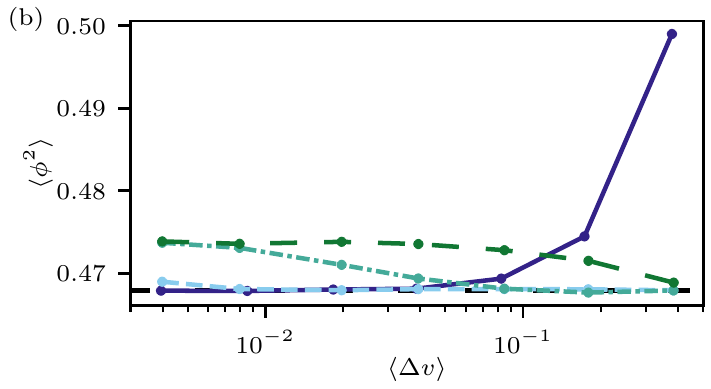}
		\hspace{0.0cm}\includegraphics{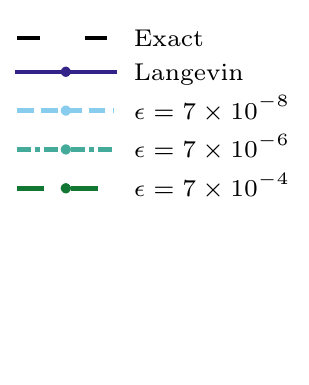}
		\includegraphics{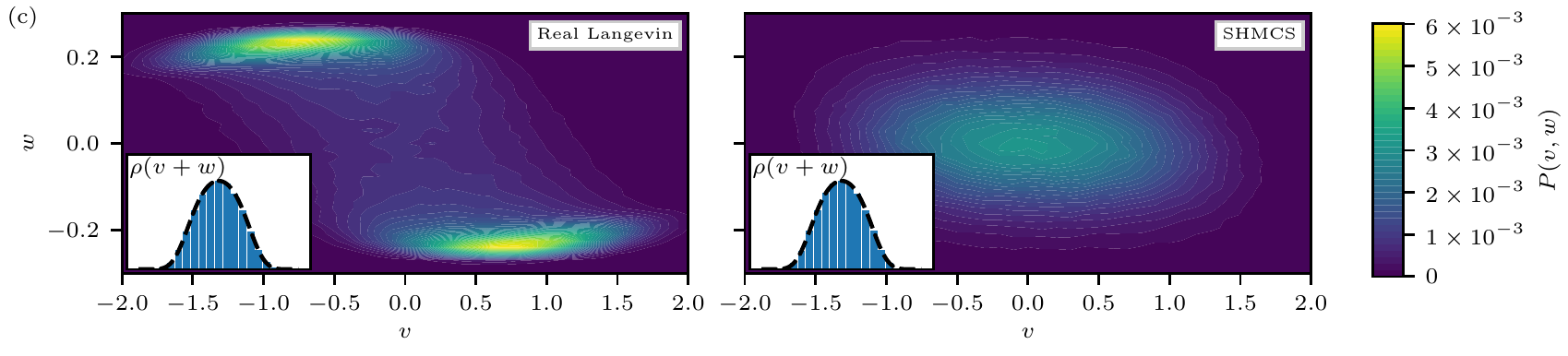}
		\caption{Comparison of the SHMCS algorithm, for $\theta=100$, and the real Langevin equation in one auxiliary dimension for the polynomial model~\eq{eq:PolynomialModel2} with $\lambda=\sigma_{\text{Re}}=1$ and $\sigma_{\text{Im}}=0$. The step size of the SHMCS algorithm in the visible direction is regulated by the evolution time with respect to the Hamilton's equations~\eq{eq:EvolutionSHMCS} and the one in the hidden direction by the parameter $\epsilon$, cf. Eq.~\eq{eq:SHMCSHiddenUpdate}. (a) Interrelationship between the step sizes in the visible and the hidden dimension. Inherent to the SHMCS algorithm, the step size in the hidden direction is independent of the real one but changes in dependence of $\epsilon$. For real Langevin dynamics, the step sizes are related to each other. (b) Convergence of the algorithms to the analytical result of the observable $\langle \phi^2 \rangle$ as a function of the step size in the visible direction. In contrast to the real Langevin equation, exact results are obtained for the SHMCs algorithm also for large visible step sizes. This is an important observation since it shows that the provided theoretical framework in this work is correct. Further, it demonstrates that, in principle, sampling from distributions with a complex contribution with larger step sizes in the visible direction is possible. The numerical results deviate for larger values of $\epsilon$ and smaller step sizes into the real direction. This property can be traced back to the constraint that the considered correlations in the visible direction need to be dominant, which is no longer the case in this limit. (c) Distribution $P(v, w)$ of the two algorithms in the higher-dimensional representation space. The histograms in the smaller plots confirm that the algorithms sample in both cases from the target distribution $\rho(\phi) = p(v + w)$, which is indicated by the dashed black line.}
		\label{fig:SHMCS}
	\end{figure*}

	The SHMCS algorithm is tested in a similar way. In this case, we compare the results with those from a real Langevin equation with one hidden variable. The algorithm is derived in the same manner as complex Langevin, but with a substitution $\phi=v + w$. Recall that the resulting Langevin equation with one hidden variable only works for real actions. We split the action according to
	\begin{equation}
	S(v + w) = S_1(v + w) + S_2(v + w)\,,
	\end{equation}
	with
	\begin{align}
	S_1(v + w) &= \frac{\sigma_{\text{Re}}}{2} w^2 + \frac{\lambda}{4} \left(v^4 + 6 v^2 w^2 + 4 v w^3\right)\,, \nonumber\\[1ex]
	S_2(v + w) &= \frac{\sigma_{\text{Re}}}{2}\left(v^2 + 2 v w\right) + \frac{\lambda}{4} \left(4 v^3 w + w^4\right)\,.
	\end{align}
	The numerical results in Fig.~\ref{fig:SHMCS} support the theoretical framework presented in this work. In particular, the SHMCS algorithm allows larger step sizes in the visible direction due to an exact satisfaction of the detailed-balance equation~\eq{eq:DetailedBalanceSubstition}. As pointed out at the end of Chapter~\ref{sec:SHMCS}, an infinitesimally small step size in the direction of the hidden variables is implemented by taking the limit $\epsilon\to 0$. The numerical results confirm the discussed restriction that stochastic contributions in the hidden variables need to be dominated by correlations in the visible variables. If the step size into the hidden direction is not small enough, compared to that in the visible direction, this domination does no longer hold. This can be observed in Fig.~\ref{fig:SHMCS} for large values of $\epsilon$ and small step sizes in the real direction.

	\begin{table*}
		\begin{tabular}{p{2.2cm}p{5.0cm}p{3.2cm}p{2.6cm}p{3.2cm}}
			Name & Transition probabilities & Proposal distribution & Action difference expansion & Real update
			\\\hline\hline
			CLE & Gaussian~(Eq.~\eq{eq:IntroComplexLangevin}) & Gaussian & 1st order & Explicit \\
			2ndCLE & Gaussian~(Eqs.~\eq{eq:CLESecondOrderReal} and~\eq{eq:CLESecondOrderImag}) & Gaussian & 2nd order & Explicit \\
			ImplGauss & Gaussian~(Eqs.~\eq{eq:ResultingTransitionComplexLangevinS} and~\eq{eq:UpdatePhiYS}) & Gaussian & Exact & Implicit \\
			MetrGauss & Gaussian~(Eqs.~\eq{eq:ResultingTransitionComplexLangevinS} and~\eq{eq:UpdatePhiYS}) & Gaussian & Exact & Metropolis\\
			ImplHat & Hat Function~(Eqs.~\eq{eq:HatImag} and~\eq{eq:HatReal}) & Hat Function & Exact & Implicit \\
			ImplUniHat & Uniform~(Eqs.~\eq{eq:UniformComplexLangevinReal} and~\eq{eq:UniformComplexLangevinImag}) & Hat Function & Exact & Implicit \\
			ImplUniUni & Uniform~(Eqs.~\eq{eq:UniformComplexLangevinReal} and~\eq{eq:UniformComplexLangevinImag}) & Uniform & Exact & Implicit \\
			MetrUniGauss & Uniform~(Eqs.~\eq{eq:UniformComplexLangevinReal} and~\eq{eq:UniformComplexLangevinImag}) & Gaussian & Exact & Metropolis \\
		\end{tabular}
		\caption{Details about the different studied algorithms in Fig.~\ref{fig:ComparisonDynamics} and Fig.~\ref{fig:RealStepSizes}. The algorithms differ in their utilized transition probabilities. For the uniform transition probability, different proposal distributions are considered. The last column indicates how the update of the real part $\phi_x$ is implemented. For complex Langevin dynamics and the second order complex Langevin algorithm, an explicit update rule can be formulated. The implicit update is performed based on a transformation of the probability density, cf. Eqs.~\eq{eq:TransformationTransitionProbabilityComplexHat} and~\eq{eq:TransformationTransitionProbability}. The Metropolis update accepts or rejects a proposal state based on Eq.~\eq{eq:MetropolisLikeSampling}.}
		\label{tab:LangevinLikeAlgorithm}
	\end{table*}

	\section{Conclusion and outlook} \label{sec:ConclusionAndOutlook}
	
	We embed complex Langevin dynamics into a generalized framework. The framework is built on the idea to substitute the integration variable in the integrals for the computation of correlations and expectation values of observables. Auxiliary parameters which are introduced by this substitution are utilized to define a Markov chain Monte Carlo algorithm that operates in a higher-dimensional state space. This space is spanned by the original representation of the state and the introduced auxiliary, hidden state variables. The sampling algorithm smoothly interpolates between different transformations of the integration variable and allows a computation of observables based on samples drawn in the Markov process. The sign problem can be circumvented in this way by a smart choice of the transition probabilities of the Markov process. Complex Langevin dynamics is derived as one possible example for such an algorithm.
	
	The introduced substitution sampling algorithm formalizes the approach as a more general algorithm that computes observables based on this idea. We provide the necessary constraints any such algorithm must be subject to. Furthermore, the algorithms derived indicate possible directions to go within the given framework.
	
	We anticipate that the presented derivation of complex Langevin dynamics provides the possibility for the development of novel algorithms and for the understanding of existing ones for simulating theories with a sign problem. For example, one might analyse a replacement of the substitution of the integral for the observables by a non-linear transformation, similar to the work in~\cite{Aarts2012} or investigate further (existing) Markov chain Monte Carlo methods in auxiliary dimensions for a possible adaptation to complex measures. Furthermore, distributions sampled by means of a process in a real extended representation space might have overlap with a distribution sampled by complex Langevin dynamics. This property makes an application of reweighting in the extended space appear attractive. The approach is similar to reweighting in the complex plane, studied in~\cite{Block2017}. Lastly, the provided mathematical constraints enable an integration of machine learning algorithms into the sampling procedure since the constraints allow the formulation of objective functions for training.
	
	\section*{Acknowledgements}
	
	We thank Felipe Attanasio, Marc Bauer, Stefanie \mbox{Czischek}, Philipp Heinen and Julian Urban for discussions. This work is supported by the Deutsche Forschungsgemeinschaft (DFG, German Research Foundation) under Germany's Excellence Strategy EXC 2181/1 - 390900948 (the Heidelberg STRUCTURES	Excellence Cluster) and under the Collaborative Research Centre SFB 1225 (ISOQUANT) and the BMBF grant 05P18VHFCA. Part of this research was performed while the author was visiting the Institute for Pure and Applied Mathematics (IPAM), which is supported by the National Science Foundation (Grant No. DMS-1440415).
	
	\appendix %
	\newpage
	
	\section{Detailed balance equation in multiple variables for different algorithms}
	\label{sec:DetailedBalanceEquationInMultipleVariablesForDifferentAlgorithms}
	
	\subsection{Hamiltonian Monte Carlo} \label{sec:HamiltonianMonteCarlo}
	
	The Hamiltonian Monte Carlo (HMC)
	algorithm assigns a momentum $\pi$ to each state $x$ of a considered system with
	probability distribution $\rho(x)\propto \exp(-S(x))$~\cite{Duane1987}. Therefore, it can be
	considered as a Monte Carlo algorithm in auxiliary dimensions. The state
	$x$ and the momentum $\pi$ can be identified with the variables $v$ and $w$. This appendix briefly demonstrates how the algorithm samples, in equilibrium, from the desired distribution $\rho(x)$. A thorough introduction to the HMC algorithm can be found, for example, in~\cite{Neal2012,Betancourt2018}.
	
	A common implementation of the HMC algorithm consists of the
	following steps:
	\begin{enumerate}
		\item Sample a momentum $\pi$ from a Gaussian distribution
		according to $q(\pi)=\varphi(\pi)$, with $\varphi(\pi)$ defined in Eq.~\eq{eq:GaussianDistribution}.
		\item Perform an integration of Hamilton's
		equations
		\begin{equation}
		\label{eq:Hamiltonsequations}
		\frac{\text{d} x}{\text{d} t} = \frac{\partial H}{\partial \pi}
		\,,\quad\, \frac{\text{d} \pi}{\text{d} t} = -\frac{\partial H}{\partial x}
		\end{equation}
		for a finite amount of time and negate the proposed momentum. We refer to the
		proposed state and momentum as
		\begin{equation}
		x' = R\,\Phi_x(x, \pi)\,,\quad\, \pi' = R\,\Phi_p(x, \pi)\,,
		\end{equation}
		where $\Phi$ represents the outcome of the integration and $R$ negates the
		resulting proposed momentum.
		\item Accept or reject the proposed state with
		probability
		\begin{equation}
		\min\left[1, \exp\left(-(H(x', \pi') - H(x,
		\pi))\right)\right]\,.
		\end{equation}
	\end{enumerate}
	
	The Hamiltonian is defined by
	\begin{equation}
	H(x,\pi) = S(x) + \frac{\pi^2}{2m}\,,
	\end{equation}
	where $S(x)$ represents the action or energy function one wants to sample from.
	
	The algorithm implements the transition probability:
	\begin{align}
	W(x, \pi& \to x', \pi') \nonumber\\[1ex]  \propto \varphi&(\pi)\,\delta\left(x' - R\,\Phi_x(x,
	\pi)\right)\,\delta\left(\pi' - R\,\Phi_\pi(x,
	\pi)\right) \nonumber\\[1ex] &\times\, \min\left[1, \exp\left(-(H(x', \pi') - H(x,
	\pi))\right)\right]\,.
	\end{align}
	The resulting equilibrium distribution in the higher-dimensional state space is given by
	\begin{equation}
	p(x, \pi) \propto \exp\left(-H(x, \pi)\right)\,.
	\end{equation}
	The Metropolis accept/reject step takes into account numerical errors in the integration scheme for $x$ and $\pi$. Otherwise, the proposal state
	can be always accepted since it holds $H(x', \pi') = H(x, \pi)$, as a result of
	Hamilton's equations.
	
	The transition probability is designed to satisfy a detailed-balance equation in the higher-dimensional state space, cf. Eq.~\eq{eq:DetailedBalanceHigherDimensions}:
	\begin{equation}
	p(x, \pi) W(x, \pi \to x', \pi') = p(x',\pi') W(x', \pi' \to x, \pi)\,.
	\end{equation}
	Detailed balance is ensured since the evolution of $x$ and $\pi$ is time-reversible and volume-preserving.
	
	Due to the statistical independence of $x$ and $\pi$, the target distribution $\rho(x)$ can be obtained by a marginalization of the joint distribution $p(x, \pi)$:
	\begin{equation}
	\rho(x) = \int\text{d}\pi\,p(x, \pi)\,.
	\end{equation}
	Hence, the sampled momenta $\pi$ can be ignored in a numerical computation of observables $\mathcal{O}(x)$.
		
	\subsection{Restricted Boltzmann machine} \label{sec:RestrictedBoltzmannMachine}
	
	Restricted Boltzmann machines (RBMs) are stochastic and generative neural networks~\cite{Smolensky1986, Freund1991, Ackley1985} typically used to parametrize probability distributions over a set of input samples. New samples can be drawn in equilibrium by Gibbs sampling. In contrast to the Hamiltonian Monte Carlo algorithm, a detailed-balance equation is only fulfilled in subsampling steps, but not for the entire update. The detailed-balance equation~\eq{eq:DetailedBalanceHigherDimensions} is not satisfied. Instead, the transition probabilities satisfy the more general constraint~\eq{eq:EquilibriumConstraint} to sample correctly from the desired equilibrium distribution, as we will show in the following. Before that, we provide a short reminder of the algorithm.
	
	The restricted Boltzmann machine consists of visible and hidden neurons, denoted as $v$ and $w$. They form a visible and a hidden layer. Each neuron can be either active or inactive. This is implemented by a binary state space, $v_i\in\lbrace 0, 1\rbrace$ and $w_j\in\lbrace 0, 1\rbrace$.
	
	An energy function can be defined in dependence of a given configuration ($v$, $w$),
	\begin{equation}
	E(v, w) = -\sum_{i} b_i v_i - \sum_{j} c_j w_j - \sum_{i, j} W_{ij} v_i w_j\,.
	\end{equation}
	The neural network parameters are given by the set $\lbrace \vec{b}, \vec{c}, \mathbf{W}\rbrace$. In contrast to the Boltzmann machine, the weight matrix is restricted to connections between single neurons of the visible and of the hidden layer. The resulting probability distributions for the RBM is defined as
	\begin{equation}
	\label{eq:EquilibriumDistributionRBM}
	p(v, w) = \frac{1}{Z} \exp\left(-E(v, w)\right)\,,
	\end{equation}
	with $Z$ being the partition sum,
	\begin{equation}
	Z=\sum_{v, w} \exp\left(-E(v, w)\right)\,.
	\end{equation}
	In general, one aims to learn a probability distribution that is defined over the visible neurons $v$. It is given by the marginal distribution
	\begin{equation}
	\rho(v) = \frac{1}{Z} \sum_{w} \exp\left(-E(v, w)\right)\,,
	\end{equation}
	where the sum runs over all possible configurations of $w$. The hidden neurons correspond to latent variables that increase the expressibility of the represented distribution.
	
	One possible approach to train the restricted Boltzmann machine is by contrastive divergence. The network parameters are adapted by a step-wise training procedure to best approximate the distribution of a training set over samples $v$. For more details, see~\cite{Ackley1985, Montfar2018}, for example.
	
	The trained restricted Boltzmann machine can be used as a generative model to draw samples from a parametrized distribution of the training set. This is realised by an update of the visible and hidden neurons based on the conditional distributions
	\begin{equation}
	\label{eq:UpdateRBMVisible}
	p(v'|w) = \frac{\exp\left(-E(v', w)\right)}{\sum_{v} \exp\left(-E(v, w)\right)}\,
	\end{equation}
	and
	\begin{equation}
	\label{eq:UpdateRBMHidden}
	p(w'|v) = \frac{\exp\left(-E(v, w')\right)}{\sum_{w} \exp\left(-E(v, w)\right)}\,.
	\end{equation}
	The sums in the denominator run again over all possible configurations of $v$ or $w$, respectively.
	
	The transition probabilities define a Markov process for the restricted Boltzmann machine. With the above definitions we are now able to analyse how the RBM samples in equilibrium from the desired distribution $p(v, w)$ by satisfying the constraint~\eq{eq:EquilibriumConstraint}, namely,
	\begin{equation}
	\label{eq:EquilibriumConstraintRBM}
	p(v', w')\stackrel{!}{=} \sum_{v, w}\, p(v, w)\, W(v, w\rightarrow v', w')\,.
	\end{equation}
	The time-dependence has been dropped since we assume the distribution to be in equilibrium.
	
	A full update step is implemented by a consecutive sampling from the conditional distributions in Eqs.~\eq{eq:UpdateRBMVisible} and \eq{eq:UpdateRBMHidden}, resulting in the transition probability
	\begin{equation}
	 W(v, w\rightarrow v', w') = p(v'|w')\, p(w'|v)\,.
	\end{equation}
	After inserting this into Eq.~\eq{eq:EquilibriumConstraintRBM}, one obtains for the right-hand side
	\begin{align}
	\sum_{v, w}\, p(&v, w)\,p(v'|w')\, p(w'|v) \nonumber\\[1ex] &= \sum_{v, w}\, p(v, w')\, p(v'|w')\, p(w|v)\,,
	\end{align}
	where we used in the second line that the transition probability $p(w'|v)$ satisfies, for a fixed $v$, the detailed-balance equation
	\begin{equation}
	p(w'|v)\,p(v, w) = p(w', w, v) = p(w|v)\,p(v, w')\,.
	\end{equation}
	We can now perform the sum over $w$ and are left with
	\begin{equation}
	p(v', w')\stackrel{!}{=}\sum_{v} p(v, w')\, p(v'|w')\,.
	\end{equation}
	After replacing $w'$ by $w$ and factoring out the transition probability, it is easy to see that the constraint matches with the transition probability of the visible variable in Eq.~\eq{eq:UpdateRBMVisible}.
	
	\section{Langevin sampling by compensation}
	\label{sec:LangevinSamplingByCompensation}
	
	In this appendix, we present a systematic approach to deriving an implementation of a substitution sampling algorithm. It represents a generalization of complex Langevin dynamics and is based on the idea to consider different proposal distributions in a respective Markov chain. We refer to the algorithm as \textit{Langevin sampling by compensation}.
	
	Our approach has two key ingredients: The first one is the reformulation of the transition probabilities of Langevin dynamics as functions of a set of visible and hidden variables. This leads to dynamics in a higher-dimensional state space. As pointed out in Chapter~\ref{sec:MarkovChainMonteCarloSamplingInAuxiliaryDimensions}, the visible and hidden variables correspond, for complex Langevin, to the real and the imaginary parts of the field.
	
	The second key ingredient is a compensation of certain contributions to the transition probabilities by terms which arise from the hidden variables. This feature is unique to the approach. In particular, it is useful for problems with a complex action where the imaginary part prevents an application of standard Monte Carlo algorithms. In these cases, the imaginary part can be compensated by the introduced imaginary part of the field. Initially complex transition probabilities can be adapted to get real-valued. The property is utilized in the complex Langevin-type algorithms in Chapter~\ref{sec:FurtherLangevinLikeAlgorithms}, which are all derived based on the here presented systematic derivation.
		
	\subsection{Complex Langevin dynamics by compensation}
	\label{sec:AppendixComplexLangevinCompensation}
	We begin with a slightly simplified derivation of complex Langevin dynamics, while using the two mentioned key ingredients. In Fig.~\ref{fig:Perspectives}, this corresponds to the transition from a real-valued action to a complex action, depicted by the golden arrow on the left-hand side. The line of arguments of this derivation is different from the standard derivation and provides a good understanding of the key ingredients. The derivation was the initial impulse for the results of this manuscript. A generalization is discussed in the next section.
	
	Consider the real Langevin
	equation~\eq{eq:LangevinDynamics}, discretized in the Langevin time $\tau$:
	\begin{equation} \label{eq:DiscreteLangevinDynamics} \phi' = \phi - \epsilon
	\frac{\delta S(\phi)}{\delta \phi} + \sqrt{2\epsilon} \eta\,,
	\end{equation}
	where $\phi' = \phi(\tau + \epsilon)$ and $\phi=\phi(\tau)$, and thus $\epsilon$ denotes a finite time step. The transition probability from state $\phi$ to $\phi'$ is given by
	\begin{equation}
	\label{eq:TransitionProbability}
	W(\phi\to\phi')=\frac{1}{\sqrt{2\epsilon}}\varphi\left(\frac{\phi'-\phi}{\sqrt{2\epsilon}}
	+  \sqrt{\frac{\epsilon}{2}}\frac{\delta S(\phi)}{\delta \phi}\right)\,,
	\end{equation} %
	where 
	\begin{equation}
	\varphi(\eta)=\frac{1}{\sqrt{2\pi}}\exp\left(-\eta^2/2\right)
	\end{equation}
	is a normalized	Gaussian distribution. $\varphi(\eta)$ is the probability with which a value $\eta$ of the noise is drawn and thus $\phi$ is updated to $\phi'$ according to the relation
	\begin{equation}
	\eta = \frac{\phi'-\phi}{\sqrt{2\epsilon}}
	+  \sqrt{\frac{\epsilon}{2}}\frac{\delta S(\phi)}{\delta \phi}\,.
	\end{equation}
	As part of the \textit{sign problem}, an accept/reject step is not possible for this transition probability if the
	action $S(\phi)$ is complex. We will show that it is
	possible to resolve this sampling problem by two mathematical tricks.
	
	In the first step, an additional variable $\phi_y'$ is introduced by choosing $\phi'$ to be complex-valued,
	\begin{equation}
	\phi' \to \phi_x' + \iu \phi_y'\,.
	\end{equation}
	The imaginary part $\phi_y'$ of the field has no physical meaning. The field $\phi'$ now lives in two
	dimensions that are spanned by its real and imaginary parts. The field
	$\phi\to\phi_x + \iu \phi_y$ will also be complex after the first update step. The argument of the Gaussian transition probability~\eq{eq:TransitionProbability} can be written, in terms of real and imaginary parts, as
	\begin{equation}
	\label{eq:ContributionTerm}
	\frac{\phi_x'+\iu\phi_y'-\phi_x - \iu \phi_y}{\sqrt{2\epsilon}}
	+  \sqrt{\frac{\epsilon}{2}}\left(\frac{\delta S_{\text{Re}}}{\delta
		\phi_x} + \iu\frac{\delta S_{\text{Im}}}{\delta \phi_x}\right)\,,
	\end{equation}
	where we define $S_{\text{Re}}:=S_{\text{Re}}(\phi_x + \iu\phi_y)$
	and $S_{\text{Im}}:=S_{\text{Im}}(\phi_x + \iu\phi_y)$. Here, we write the functional derivative of $S(\phi_x + \iu \phi_y)$ with respect to the physical field variable $\phi_x$ since our initial field $\phi$
	is identified with $\phi_x$, whereas $\phi_y$
	represents an additional variable. This in concordance with the introduction of a complex field in Sec.~\ref{sec:KeyFindings} and Sec.~\ref{sec:ComplexLangevinVersusHMCRBM}.
	
	The second important step is to choose the free variable $\phi_y'$ in such a way that it
	compensates all imaginary contributions in the argument~\eq{eq:ContributionTerm} in the transition probability, which arises from the imaginary part of the action. This is accomplished by setting
	\begin{equation}
	\label{eq:UpdateY} \phi_y' = \phi_y - \epsilon
	\frac{\delta S_{\text{Im}}}{\delta \phi_x}\,.
	\end{equation}
	As a result of this, despite a complex
	action $S$, the function $\varphi$ in the transition probability has a real argument and thus represents a valid probability distribution for the Langevin update, cf. Eq.~\eq{eq:TransitionProbability}. Sampling from this distribution is achieved by the Langevin update rule~\eq{eq:DiscreteLangevinDynamics} for $\phi_x$,
	\begin{equation}
	\label{eq:UpdateX}
	\phi_x' = \phi_x - \epsilon
	\frac{\delta S_{\text{Re}}}{\delta \phi_x} + \sqrt{2\epsilon} \eta\,.
	\end{equation}
	The update equations~\eq{eq:UpdateY} and~\eq{eq:UpdateX} are equivalent to the discretized update rules of complex Langevin dynamics~\eq{eq:IntroComplexLangevin} since, for holomorphic actions,
	\begin{align}
	\label{eq:Identifications}
	\frac{\delta S_{\text{Re}}}{\delta
		\phi_x} &= \text{Re}\left[\frac{\delta S}{\delta \phi}\bigg|_{\phi_x + \iu
		\phi_y}\right]\,,\nonumber\\[1ex] \frac{\delta S_{\text{Im}}}{\delta \phi_x} &=
	\text{Im}\left[\frac{\delta S}{\delta \phi}\bigg|_{\phi_x
		+ \iu \phi_y}\right]\,.
	\end{align}
	\subsection{Systematic derivation}
	\label{sec:SystematicDerivation}
	
	We continue with a generalization of this derivation that permits the usage of different kinds of proposal distributions. The systematic step-by-step approach provides transition probabilities $T$ and $g$ for the visible and hidden variables. These are constructed to satisfy the constraints of a substitution algorithm that were pointed out at the beginning of Chapter~\ref{sec:SubstituionSampling}. This is achieved by aiming at an implementation of the Langevin symmetry~\eq{eq:SymmetryLangevin} and of the adapted detailed-balance equation~\eq{eq:DetailedBalanceT}.
	
	The systematic approach starts with the definition of a standard Monte Carlo algorithm and continues with an application to a complex action. Accordingly, the derivation follows, in contrast to the previous derivation, the directions of the red arrows in Fig.~\ref{fig:Perspectives}.
	
	Similar to complex Langevin dynamics, the necessary constraints are fulfilled, by definition, only in the limit of infinitesimally small step sizes in configuration space. A reliable estimation of observables is only feasible for an extrapolation to infinitesimally small step sizes.
	
	\begin{figure*}
	\centering \includegraphics[width=\linewidth]{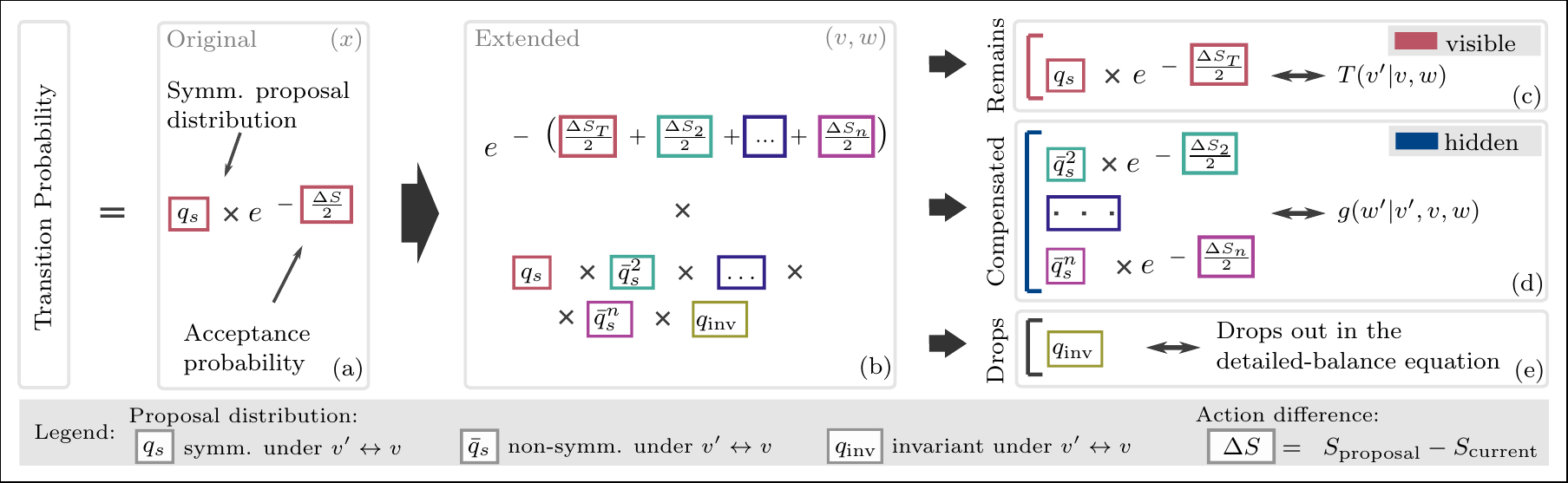}
	\caption{Step by step illustration of the systematic derivation of a Langevin sampling by compensation algorithm in App.~\ref{sec:SystematicDerivation}. We consider an initial transition probability in the original representation space that can be written as product of a symmetric proposal distribution for $x'$ and an acceptance probability that depends on the change in the action, cf. Eqs.~\eq{eq:SymmetricProposalDistribution} and~\eq{eq:AcceptanceProbability}. After a transition to the extended representation space defined in $(v, w)$ (see Eq.~\eq{eq:MappedDistributions}), symmetric, non-symmetric and invariant parts of the proposal distribution can be extracted. The scheme illustrates the case where the proposal distribution can be decomposed into a product of symmetric and non-symmetric terms, cf. Eq.~\eq{eq:DecompositionProposalDistribution}. This kind of decomposition is also used for a derivation of complex Langevin dynamics. In the next step, different action contributions, defined in Eq.~\eq{eq:ActionDecomposition}, are assigned to the different terms of the proposal distribution. This matching allows a definition of the transition probabilities for the visible and the hidden variables as illustrated on the right-hand side. The update of the hidden variables is based on the idea to utilise the updated hidden variables $w'$ to compensate the associated action contributions, cf. Eqs.~\eq{eq:HiddenVariableConstraint} and~\eq{eq:HiddenVariableConstraint2}. Furthermore, the invariant term in the proposal distribution drops out in the adapted detailed-balance equation~\eq{eq:DetailedBalanceLangevin2}.} \label{fig:SchematicSystemDerivation}
	\end{figure*}
	
	For comparison, we state, in parallel to the general approach, the specific equations for the case of complex Langevin. The different steps of the derivation are sketched, for the case of complex Langevin, in Fig.~\ref{fig:SchematicSystemDerivation}. For the more general derivation, we keep the notation in terms of visible and hidden variables. For complex Langevin dynamics these correspond to the real and imaginary parts of the field. The field $\phi$ is represented by $x$.
	
	Our derivation consists of the following steps: We start with a given proposal distribution and acceptance probability for a Markov chain Monte Carlo algorithm. Next, the representation of the state as well as these distributions are extended by auxiliary dimensions based on the substitution $x = v + w$. Following the provided theoretical framework for the substitution sampling algorithm, the dynamics is extended to take place in both the visible variables $v$ and the hidden variables $w$. This introduces the constraints no.~\ref{item:1} to no.~\ref{item:4} on the algorithm to be taken into account, as defined in Sec.~\ref{sec:GeneralDefinition}. An identification of symmetric and non-symmetric terms with respect to an exchange of $v'$ and $v$ will allow defining transition probabilities $g(w'|v',v,w)$ that satisfy the Langevin symmetry~\eq{eq:SymmetryLangevin}.
	
	\subsubsection*{Setting up a Markov chain Monte Carlo algorithm}
	
	In a Markov chain Monte Carlo (MCMC) algorithm, a new state $x'$ is proposed according to a
	distribution $q(x\to x')$ for a given state $x$. We restrict ourselves to
	symmetric proposal distributions with
	\begin{equation}
	\label{eq:SymmetricProposalDistribution}
	q(x\to x') = q(x' \to x)\,.
	\end{equation}
	It will turn out that the adapted detailed-balance equation~\eq{eq:DetailedBalanceT} can be satisfied for this algorithm only in the limit of infinitesimally small differences between the proposed state and the current state. The proposal distribution is constrained by this restriction. Hence, representations of the delta-distribution are applicable proposal distributions under these conditions. Recall that for complex Langevin dynamics, the proposal distribution is a Gaussian distribution,
	\begin{equation}
	\label{eq:GaussianProposalDistribution}
	q(\phi \to \phi')  = \frac{1}{\sqrt{2\epsilon}}\varphi\left(\frac{\phi'-\phi}{\sqrt{2\epsilon}}\right)\,.
	\end{equation}
	The transition probability $W(x\to x')$ for $x\to x'$ is commonly expressed as a product of the proposal
	probability $q(x\to x')$ and an acceptance probability $A(x\to x')$,
	\begin{equation}
	W(x \to x') = q(x \to x')\,A(x \to x')\,,
	\end{equation}
	see step~(a) in Fig.~\ref{fig:SchematicSystemDerivation}. If the acceptance probability is written in the exponential form
	\begin{equation}
	\label{eq:AcceptanceProbability}
	A(x\to x') \propto \exp\left(-\frac{\Delta S(x', x),}{2}\right)\,,
	\end{equation}
	with $\Delta S(x', x) = S(x') - S(x)$, the resulting transition probability $W$ satisfies the detailed-balance equation~\eq{eq:DetailedBalance}, with $\rho(x) = Z^{-1}\exp\left(-S(x)\right)$. This is the standard procedure in any Metropolis-Hastings algorithm.
	
	\subsubsection*{Extending the representation space}
	
	In the following we extend the procedure to
	a higher-dimensional representation space. This is achieved by the substitution $x = v + w$, where the higher-dimensional space is spanned by the set $(v, w)$ of visible and hidden variables and where $v$ has the same dimension as $x$. The purpose of this is to use the state variables in the resulting auxiliary dimensions to compensate certain contributions of the action, as shown below. For the example of a complex action, we define $\phi = \phi_x + \iu \phi_y$ and aim to compensate the imaginary contribution of the action by the imaginary part of the field.
	
	Next, we replace $x$ in the steady-state distribution and in the
	transition probability by its higher-dimensional representation
	\begin{align}
	\label{eq:MappedDistributions}
	x&\to
	v + w\,,\nonumber\\[1ex] \rho(x)&\to \rho(v + w)=:p(v, w)\,,\nonumber\\[1ex] q(x\to x')&\to q(v, w\to v', w')\,,\nonumber\\[1ex] A(x\to x')&\to
	A(v, w\to v', w')\,,
	\end{align}
	as is also indicated in step~(b) in Fig.~\ref{fig:SchematicSystemDerivation}. The resulting distributions in general do not satisfy the constraints a substitution algorithm is subject to. It may, in practice, be impossible to sample from a given proposal distribution and to evaluate the acceptance probability of a proposed state. This is, for example, the case for complex Langevin dynamics, where the action is complex and thus $w$ becomes imaginary. Accordingly, all the distributions are complex and represent no longer probability distributions. However, as in the special case of complex Langevin dynamics, there is a way around these problems that allows defining transition probabilities $g(w'|v',v,w)$
	and $T(v'|v, w)$.
	
	\subsubsection*{The acceptance probability}
	
	We start by considering the acceptance probability in the higher-dimensional space. Based on the substitutions in Eq.~\eq{eq:MappedDistributions}, it determines the likelihood of a
	proposed state $(v', w')$. This implies a
	change in both, $v$ and $w$. However, we
	aim to define a transition probability $T(v'|v, w)$ that ensures that in the long-time limit the adapted detailed-balance equation~\eq{eq:DetailedBalanceT} is fulfilled, by satisfying
	\begin{equation}
	\label{eq:DetailedBalanceLangevin2}
	p(v,w)\,T(v'|v, w) =
	p(v',w)\,T(v|v', w)\,.
	\end{equation}
	Note that the steady-state distribution $p$ is
	evaluated on both sides of the equation at the same hidden state $w$. Hence, also the acceptance probability needs to account for changes in $v$ only. Therefore, we define
	\begin{equation}
	\label{eq:LangevinAcceptanceTerm} A(v\to v'|w) \propto
	\exp\left(-\frac{S(v',w) -
		S(v, w)}{2}\right)\,,
	\end{equation}
	where $S(v, w):= S(v + w)$. This choice reflects the property of the substitution sampling algorithm to incorporate a (dominant) stochastic contribution only into the direction of the visible variables, cf. Sec.~\ref{sec:Construction}. In the case of complex Langevin dynamics, the imaginary part $\phi_y$ of the field is kept constant and a change of the real part reflects the expectation value with respect to the original field $\phi$.
	
\begin{figure}
		\includegraphics[width=\linewidth]{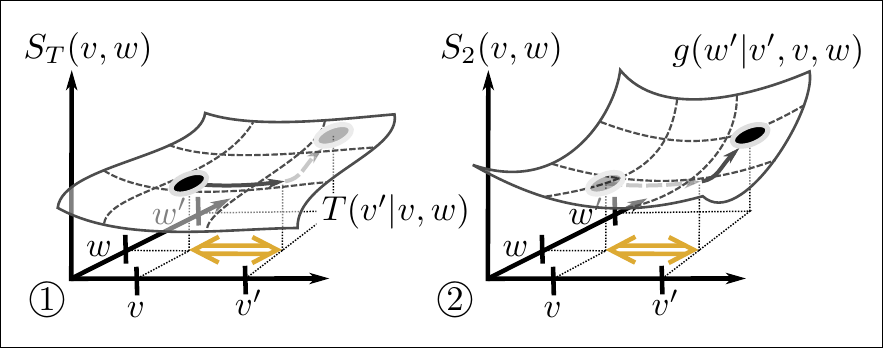}
		\caption{Dependence of the action on the transition probabilities for the Langevin sampling by compensation algorithm. Only the first part of the update step on the left-hand side is stochastic. The transition probability $T(v'|v, w)$ depends on the action difference $S_T(v', w) - S_T(v, w)$. In contrast, the second part of the update step is deterministic. The updated hidden variable $w'$ is determined by $g(w'|v',v, w)$ and depends on the action difference $S_2(v', w) - S_2(v, w)$. It is important to note that in both cases $w$ is fixed and the action difference is calculated with respect to a change in the visible variable $v$. This is emphasized by the golden double arrow in both illustrations.}
		\label{fig:Dynamics}
	\end{figure}

	We want to construct the transition probability $T$ in $v$ as a product of a proposal distribution and
	an acceptance probability. The acceptance
	probability~\eq{eq:LangevinAcceptanceTerm} already satisfies the adapted detailed-balance
	equation~\eq{eq:DetailedBalanceLangevin2} for a given transition
	$v\to v'$ as long as (i)~the proposal distribution is symmetric under an
	exchange of $v'$ and $v$ and (ii)~the transition probabilities refer to the same hidden variable $w$ as starting point for the next update. The latter condition is depicted in Fig.~\ref{fig:Symmetry} and by the golden double arrow in the first part of the update step in Fig.~\ref{fig:Dynamics}.
	
	\subsubsection*{Symmetries}
	
	A suitable proposal distribution for $T$ as well as a definition for the transition probability $g$ are derived in the following by a distinction of symmetric and non-symmetric terms in the higher-dimensional distributions in Eq.~\eq{eq:MappedDistributions}. The procedure is also sketched in part (b) in Fig.~\ref{fig:SchematicSystemDerivation}.
	
	We distinguish different terms in the proposal distribution
	$q(v, w\to v', w')$. Terms that are symmetric under an exchange of $v'$ and $v$ are denoted as $q_s$ whereas non-symmetric terms are referred to as $\bar{q}_s$. Factors that do not depend on the visible variables are denoted as $q_\text{inv}$. The actual relation between these terms depends on the proposal distribution. For example, the terms form a product for a Gaussian proposal distribution,
	\begin{equation}
	\label{eq:DecompositionProposalDistribution}
	q(v, w\to v', w') = q_s \times \bar{q}_s^2 \times \cdots \times \bar{q}_s^n \times q_\text{inv.}\,.
	\end{equation}
	For complex Langevin dynamics, the following factors can be identified in the proposal distribution~\eq{eq:GaussianProposalDistribution} after a substitution of $\phi$ by $\phi_x + \iu \phi_y$:
	\begin{align}
	\label{eq:DecomposedProposalDistribution}
	q_s & = \frac{1}{\sqrt{4\pi\epsilon}}\exp\left(-\frac{(\phi_x'-\phi_x)^2}{2\epsilon}\right)\,,\nonumber\\[1ex]
	\bar{q}_s^2 &= \exp\left(-\frac{\iu}{2}\left[\frac{(\phi_x'-\phi_x)(\phi_y'-\phi_y)}{\epsilon}\right]\right)\,,\nonumber\\[1ex]
	q_\text{inv} &= \exp\left(\frac{(\phi_y'-\phi_y)^2}{2\epsilon}\right)\,.
	\end{align}
	In the case of the complex hat function algorithm in App.~\ref{sec:AppendixComplexHatFunctionAlgorithm}, the proposal distribution can be expressed as a sum
	\begin{equation}
	\label{eq:DecompositionProposalDistributionSum}
	q(v, w\to v', w') = q_s + \bar{q}_s\,.
	\end{equation}
	The total number of terms depends on the number of auxiliary variables.
	
	Recall that we want to define the transition probability as a product of a proposal distribution and the acceptance probability~\eq{eq:LangevinAcceptanceTerm}. Keeping this in mind, the following findings are an important result of the above distinction.
	
	On the one hand, the non-symmetric terms need to vanish in the proposal distribution for a fulfilment of the detailed-balance equation~\eq{eq:DetailedBalanceLangevin2}, at least in the statistical mean. On the other hand, we want to compensate certain contributions, such as the imaginary ones in the case of complex Langevin dynamics, in the action difference in the acceptance probability~\eq{eq:LangevinAcceptanceTerm} that make it otherwise infeasible to sample. This is the main motivation of the entire approach.
	
	\subsubsection*{Deriving $T(v'|v, w)$}
	
	We prepare the desired compensation of certain action terms by a decomposition of the acceptance probability into symmetric and non-symmetric terms and by matching these with the terms of the proposal distribution. First, we decompose the action $S(v, w)$ into $n$ terms,
	\begin{align}
	\label{eq:ActionDecomposition}
	S(&v,
	w)\nonumber\\[1ex] &=S_T(v, w) +
	S_2(v, w) + \ldots + S_n(v, w)\,,
	\end{align}
	where $S_T$ is used to the define the transition probability $T$. The terms $S_2, \ldots, S_n$ will be compensated by use of the hidden variables $w'$.
	
	For a complex action the above corresponds to a separation of the real and imaginary parts. We define, for this case,
	\begin{align}
	S_T(\phi_x,\phi_y)&=S_{\text{Re}}(\phi_x + \iu\phi_y)\,,\nonumber\\[1ex] S_2(\phi_x,\phi_y)&=\iu S_{\text{Im}}(\phi_x + \iu\phi_y)\,.
	\end{align}
	We associate the real part of the action with the update of $\phi_x$ and the imaginary part with that of $\phi_y$.
	
	Next, we analogously decompose the acceptance probability. The actual decomposition is dictated by the form of the proposal distribution. In the case of the product~\eq{eq:DecompositionProposalDistribution}, one defines
	\begin{align}
	\label{eq:DecomposedAcceptanceProbability}
	A(v\to v'|w) \propto \exp&\left(-\frac{S_T(v',
	w) - S_T(v, w)}{2}\right) \nonumber\\[1ex] & \;\qquad\qquad\times \bar{A}_s^2\times \cdots \times \bar{A}_s^n\,.
	\end{align}
	For a sum, such as Eq.~\eq{eq:DecompositionProposalDistributionSum}, a possible decomposition is
	\begin{align}
	A(v\to v'|w)& \propto \exp\left(-\frac{S_T(v',
	w) - S_T(v, w)}{2}\right) \nonumber\\[1ex] &\quad\;\;\times \left[A_s^2 + \bar{A}_s^2 + \ldots + A_s^n + \bar{A}_s^n\right]\,.
	\end{align}
	As derived before, a change in the action is only considered in the visible direction. At this point, it is sufficient to focus on the symmetric and non-symmetric terms, $A_s^i$ and $\bar{A}_s^i$. This allows, in the following step, a definition of $T$ and $g$.
	
	For our example of a complex action, the mathematical operation is a product and the non-symmetric term $\bar{A}_s^2$ is given by
	\begin{equation}
	\bar{A}_s^2 = \exp\left(-\iu\,\frac{S_\text{Im}(\phi_x' + \iu\phi_y) - S_\text{Im}(\phi_x + \iu\phi_y)}{2}\right)\,.
	\end{equation}
	We continue by considering the product of the decomposed proposal and acceptance probabilities, namely:
	\begin{equation}
	\label{eq:ComposedProduct}
	q(v, w\to v', w') \times A(v\to v'|w)\,,
	\end{equation}
	collecting all terms symmetric with respect to an exchange of $v'$ and $v$, to define the transition probability $T$. For example, for the product form~\eq{eq:DecompositionProposalDistribution}, the transition probability is defined as
	\begin{equation}
	\label{eq:FinalTransitionProbability}
	T(v'|v, w) \propto q_s \times \exp\left(-\frac{S_T(v', w) -
		S_T(v, w)}{2}\right)\,.
	\end{equation}
	The right-hand side consists of a product of the symmetric term $q_s$ in Eq.~\eq{eq:DecompositionProposalDistribution} and of the first factor of the acceptance probability in Eq.~\eq{eq:DecomposedAcceptanceProbability}. For the example of complex Langevin dynamics, this combination of symmetric terms is shown in step~(c) in Fig.~\ref{fig:SchematicSystemDerivation}. In this case, the transition probability for the real part of the field, with the Gaussian $q_s\sim \varphi$, reads
	\begin{align}
	\label{eq:ResultingTransitionComplexLangevin}
	T(\phi_x'&|\phi_x, \phi_y) \nonumber\\[1ex] \propto&\,\frac{1}{\sqrt{2\epsilon}}\varphi\left(\frac{\phi_x'-\phi_x}{\sqrt{2\epsilon}}\right)\exp\left(-\frac{\Delta S_{\text{Re}}(\phi', \phi)}{2}\right)\,,
	\end{align}
	where $\Delta S_\text{Re}(\phi', \phi) = S_\text{Re}(\phi_x' + \iu\phi_y) - S_\text{Re}(\phi_x + \iu\phi_y)$.
	
	We will study, in App.~\ref{sec:Implications}, under which conditions the transition probability satisfies the adapted detailed-balance equation~\eq{eq:DetailedBalanceLangevin2}.
	
	\subsubsection*{Deriving $g(w'|v', v, w)$}
	
	It remains to determine a transition probability $g(w'|v', v, w)$, which satisfies the Langevin symmetry~\eq{eq:SymmetryLangevin},
	\begin{equation}
	\label{eq:SymmetryLangevin2}
	g(w' | v', v,
	w) \stackrel{!}{=} g(w' | v, v', w)\,,
	\end{equation}
	as suggested for a substitution sampling algorithm, cf. Sec.~\ref{sec:Construction}.
	
	The above distinction of symmetric and non-symmetric terms allows determining the transition probability $g$ by compensating the remaining terms in the above discussed product of the proposal distribution and the acceptance probability. More specifically, all terms of the product~\eq{eq:ComposedProduct} that do not contribute to the transition probability~\eq{eq:FinalTransitionProbability} are supposed to cancel each other, for which we will use the updated hidden variables $w'$.
	
	Considering first again the case of the product form~\eq{eq:DecompositionProposalDistribution} of the proposal distribution, this translates into
	\begin{align}
	\label{eq:HiddenVariableConstraint}
	\bar{q}_s^2 \times &\cdots \times \bar{q}_s^n \times \bar{A}_s^2 \times \cdots \times \bar{A}_s^n \stackrel{!}{=} 1\nonumber\\[1ex]&\Leftrightarrow\quad w' - h(v', v, w) \stackrel{!}{=} 0\,,
	\end{align}
	with
	\begin{equation}
	\label{eq:HiddenVariableConstraint2}
	g(w'|v', v, w) = \delta\left(w' - h(v', w, w)\right)\,.
	\end{equation}
	The matching of the remaining terms is illustrated in step~(d) in Fig.~\ref{fig:SchematicSystemDerivation}. Following Sec.~\ref{sec:Construction}, the function $h(v', v, w)$ defines the updated value of $w'$. The invariant term $q_\text{inv}$ has been neglected as it can be cancelled in the adapted detailed-balance equation.
	
	This is always possible since the updated state $w'$ can be chosen arbitrarily as long as the update rule satisfies the Langevin symmetry. As a result of the symmetric properties of the remaining terms, the resulting transition probability indeed bears this symmetry.
	
	In the case of complex Langevin, Eq.~\eq{eq:HiddenVariableConstraint} can be simplified to
	\begin{equation}
	\frac{(\phi_x'-\phi_x)(\phi_y'-\phi_y)}{\epsilon} + \Delta
	S_{\text{Im}}(\phi', \phi)\,\stackrel{!}{=}0\,,
	\end{equation}
	with $\Delta S_\text{Im}(\phi', \phi) = S_\text{Im}(\phi_x' + \iu\phi_y) - S_\text{Im}(\phi_x + \iu\phi_y)$. Consequently, the update rule for the hidden state is
	\begin{equation}
	\label{eq:UpdatePhiY}
	\phi_y' = \phi_y - \epsilon	\frac{\Delta S_{\text{Im}}(\phi', \phi)}{\phi_x' - \phi_x}\,.
	\end{equation}
	As intended, the updated imaginary part of the field compensates imaginary contributions arising in the product~\eq{eq:ComposedProduct} of the proposal distribution and the acceptance probability. The compensation has the same effect as in complex Langevin dynamics in the previous section, namely, resulting in a real-valued transition probability $T(\phi_x'|\phi_x, \phi_y)$ for the real part of the field.
	
	The compensation is either exact or satisfied in a stochastic way through $h(v', v, w)$. An example for a stochastic update of the hidden variable $w$ is given by complex Langevin with imaginary noise, see, for example,~\cite{Aarts2010}. Thereby, it is however important that the stochastic behaviour in the visible direction is dominant. This restriction is reflected by the constraints on a substitution sampling algorithm, defined in Sec.~\ref{sec:GeneralDefinition}.
	
	\subsection{Implications}
	\label{sec:Implications}
	
	The derived transition probabilities do not yet satisfy all of the constraints a substitution sampling algorithm is subject to. In the following, we derive further restrictions which ensure this, analogous to the discussion for complex Langevin in Sec.~\ref{sec:ComplexLangevinAsASubstitutionSamplingAlgorithm}.
	
	The adapted detailed-balance equation~\eq{eq:DetailedBalanceLangevin2} is violated for the transition probabilities $T(v'|v, w)$, resulting in
	\begin{align}
	\label{eq:DetailedBalanceAdapted}
	&p(v, w) T(v' |v, w) \nonumber\\[1ex] &= p(v', w)  T(v | v', w)\exp\left(-\sum_{i=2}^n (S_i(v, w) - S_i(v', w))\right)\,,
	\end{align}
	where $p(v, w) = \rho(v + w)$. This is the same discrepancy as for the transition probability $T(\phi_x'|\phi_x, \phi_y)$ of complex Langevin in Sec.~\ref{sec:ComplexLangevinAsASubstitutionSamplingAlgorithm}, cf. Eq.~\eq{eq:DetailedBalanceLangevinAdapted}. It can be traced back to the restriction to the terms $S_T$ in the action difference, cf. Eqs.~\eq{eq:ActionDecomposition} and~\eq{eq:FinalTransitionProbability}.
	
	The discrepancy can be resolved by imposing
	\begin{equation}
	\label{eq:LangevinViolation}
	\exp\left(-\sum_{i=2}^n (S_i(v, w) - S_i(v', w))\right) \stackrel{!}{=} 1\,.
	\end{equation}
	This can be reached with infinitesimal stepping in updating the visible variable $v$. The proposal distribution needs to allow implementing this limit. As pointed out previously, representations of delta-distributions are examples for appropriate proposal distributions. Since an infinitesimally small sampling step is not meaningful algorithmically, we resort to an extrapolation towards zero step size.
	
	We conclude that the restriction to an infinitesimal step size in the visible direction entails a satisfaction of the adapted detailed-balance equation, cf. Eqs.~\eq{eq:DetailedBalanceT} and~\eq{eq:DetailedBalanceLangevin2}. Recalling that the transition probability $g$ of the hidden variables implements the Langevin symmetry by construction, we find that constraint no.~\ref{item:1} for a substitution sampling algorithm	is fulfilled.
	
	Constraint no.~\ref{item:2} requires that the step size in the direction of the hidden variables is infinitesimal. As the transition probabilities $T$ and $g$ are derived from the same proposal distribution, the step size of the hidden variables is already reduced simultaneously with the one in the visible direction.

	Constraint no.~\ref{item:3} depends on the considered model.
	
	It remains an analysis of constraint no.~\ref{item:4}. It is not possible to show that this is generally fulfilled for arbitrary proposal distributions. For the case of complex Langevin dynamics it is proven in Sec.~\ref{sec:ComplexLangevinAsASubstitutionSamplingAlgorithm}. We assume that the proof is also valid for other proposal distributions as long as these coincide in the limit of infinitesimally small step sizes with a delta-distribution. This assumption is supported by the numerical results in Chapter~\ref{sec:NumericalResults}.
	
	Keeping this in mind, the restrictions on $g$, in constructing substitution sampling algorithms, cf. Eq.~\eq{eq:HiddenTransitionConstraint2}, can be relaxed to
	\begin{equation}
	\label{eq:HiddenTransitionConstraint3}
	g(w'|v', v, w) = \delta\left(w' - h(v', v, w; \epsilon)\right)\,,
	\end{equation}
	where the parameter $\epsilon$ and the function $h$ ensure, as before, an infinitesimal step size in the hidden direction:
	\begin{equation}
	\lim\limits_{\epsilon\to 0} h(v', v, w; \epsilon) = w\,.
	\end{equation}
	More specifically, we reinserted a dependence of the transition probability on the updated visible state $v'$. This relaxation is, for example, utilized in the complex hat function algorithm in Sec.~\ref{sec:ComplexHatFunctionAlgorithm}.
	
	The derivation of the discretized update equations for complex Langevin dynamics is completed in App.~\ref{sec:ComplexLangevinDynamicsCLELike}.
	
	\subsection{Measure for accuracy} \label{sec:MeausreForAccuracy}
	
	In the previous section, it has been shown that the detailed-balance
	equation is violated for simulations with a finite step size in the visible
	states $v$. One can define a measure $\kappa$ for the accuracy of the Langevin sampling by compensation algorithm based on Eq.~\eq{eq:LangevinViolation}, %
	\begin{equation}
	\label{eq:AccuracyMeasure}
	\kappa(v', v, w) = \bigg|\sum_{i=2}^n S_i(v', w) - \sum_{i=2}^n S_i(v, w)\bigg|\,.
	\end{equation}
	It measures the violation of the detailed-balance equation in dependence on the step size in $v$. A simulation satisfies the detailed-balance equation if $\kappa(v', v, w) = 0$. Our numerical results in Chapter~\ref{sec:NumericalResults} confirm that
	$\kappa$ represents a reasonable measure in analysing the Langevin sampling by compensation algorithm for finite step sizes.
	
	The measure is in accordance with an improved numerical stability of complex Langevin dynamics by introducing an adaptive step size~\cite{Aarts2008, Aarts20102}. This adaptation prevents too large step sizes, leading to small measures of $\kappa(v', v, w)$.
	
	\section{Complex Langevin-type sampling by compensation algorithms}
	\label{sec:CLELikeAlgorithms}
	
	\subsection{Complex Langevin dynamics}
	\label{sec:ComplexLangevinDynamicsCLELike}

	For completeness, the discretized update equations of complex Langevin dynamics in Eq.~\eq{eq:IntroComplexLangevin} are derived explicitly from the transition probabilities for complex actions, defined in App.~\ref{sec:SystematicDerivation}:
	\begin{align}
	\label{eq:ResultingTransitionComplexLangevin2}
	T(\phi_x'&|\phi_x, \phi_y)\nonumber\\[1ex] \propto&\,\frac{1}{\sqrt{2\epsilon}}\varphi\left(\frac{\phi_x'-\phi_x}{\sqrt{2\epsilon}}\right)\exp\left(-\frac{\Delta S_{\text{Re}}(\phi', \phi)}{2}\right)\,,
	\end{align}
	and
	\begin{equation}
	\label{eq:UpdatePhiY2}
	\phi_y' = \phi_y - \epsilon	\frac{\Delta S_{\text{Im}}(\phi', \phi)}{\phi_x' - \phi_x}\,.
	\end{equation}
	The action difference $\Delta S(\phi', \phi)$ can be expanded
	around $\phi_x$ since the step sizes in the real direction are constraint to be infinitesimally small:
	\begin{align}
	\label{eq:Taylor}
	\Delta S&(\phi', \phi) = S(\phi_x +
	\delta \phi_x + \iu\phi_y) - S(\phi_x + \iu\phi_y) \nonumber\\[1ex]&
	\simeq \delta \phi_x \frac{\delta S(\phi_x + \iu\phi_y)}{\delta \phi_x} = \delta \phi_x \left[\frac{\delta S(\phi)}{\delta \phi}\bigg|_{\phi_x + \iu \phi_y}\right]\,,
	\end{align}
	where
	\begin{align}
	S(\phi_x + \iu \phi_y) & = S_\text{Re}(\phi_x +\iu\phi_y) + \iu S_\text{Im}(\phi_x+\iu \phi_y) \nonumber\\[1ex]
	& \equiv S_\text{Re} + \iu S_\text{Im}\,.
	\end{align}
	
	The expansion simplifies the update rule~\eq{eq:UpdatePhiY2} of the imaginary part $\phi_y$, resulting in
	\begin{equation}
	\phi_y' = \phi_y - \epsilon\frac{\delta S_\text{Im}}{\delta \phi_x}\,,
	\end{equation}
	the discrete update
	dynamics of the imaginary field $\phi_y$ in complex Langevin dynamics.
	
	The transition
	probability~\eq{eq:ResultingTransitionComplexLangevin2} turns, with this expansion, into
	\begin{equation}
	\label{eq:TransitionProbabilityReal}
	T(\phi_x' | \phi_x, \phi_y) =\frac{1}{\sqrt{2\epsilon}}\varphi\left(\frac{\phi_x'-\phi_x}{\sqrt{2\epsilon}} +  \sqrt{\frac{\epsilon}{2}}\frac{\delta S_\text{Re}}{\delta \phi_x} \right)\,.
	\end{equation}
	The absorption of the action term into the Gaussian distribution can be shown by expanding the argument
	of the exponential function in Eq.~\eq{eq:ResultingTransitionComplexLangevin2} with a
	first order term in $\epsilon$. The first Gaussian distribution
	and the exponential term are contracted by completing the square in the exponent. The explicit computation can be found in App.~A of reference~\cite{Kades2020} and is similar to the computation in App.~\ref{sec:AbsorbingTheImaginaryContribution}.
	
	An explicit update rule for $\phi_x$ can be derived by a transformation of the transition probability, by demanding
	\begin{equation}
	\label{eq:TransformationTransitionProbability}
	\int_{-\infty}^{\phi_x'} \text{d}\tilde{\phi}_x \, T(\tilde{\phi}_x | \phi_x, \phi_y)
	\stackrel{!}{=} \int_{-\infty}^\eta \text{d}\tilde{\eta} \,
	\varphi(\tilde{\eta})\,.
	\end{equation}
	Evaluating both integrals and solving for $\phi_x'$ results in the discrete update rule:
	\begin{equation}
	\phi_x' = \phi_x - \epsilon\frac{\delta S_\text{Re}}{\delta \phi_x} + \sqrt{2\epsilon} \eta\,.
	\end{equation}
	By using the relations in Eq.~\eq{eq:Identifications}, the derived update rules coincide with the ones of complex Langevin dynamics, Eq.~\eq{eq:IntroComplexLangevin}.
		
	\subsection{Second order complex Langevin}
	\label{sec:AppendixSecondOrderComplexLangevin}
	
	It is possible to formulate a discrete second order complex Langevin equation.
	The derivation follows the same line of argumentation as in the previous section, and the name refers to the second order terms in the expansion of the action
	difference for infinitesimally small step sizes. We keep this term in the
	expansion in Eq.~\eq{eq:Taylor},
	\begin{align}
	\label{eq:SecondOrderExpansion}
	\Delta S&(\phi', \phi) = S(\phi_x + \delta \phi_x + \iu
	\phi_y) - S(\phi_x + \iu\phi_y) \nonumber\\[1ex]&\simeq \delta \phi_x
	\frac{\delta S(\phi_x + \iu\phi_y)}{\delta \phi_x} + \frac{\delta
		\phi_x^2}{2} \frac{\delta^2 S(\phi_x + \iu \phi_y)}{\delta \phi_x^2}\,.
	\end{align}
	The second order expansion of the action difference in the imaginary part can be inserted into Eq.~\eq{eq:UpdatePhiY2}, the update rule of the imaginary field $\phi_y$. This results in 
	\begin{equation}
	\phi_y' = \phi_y - \epsilon \frac{\delta
		S_\text{Im}}{\delta \phi_x} - \frac{\epsilon}{2}\left(\phi_x' - \phi_x\right) \frac{\delta^2
		S_\text{Im}}{\delta \phi_x^2}\,,
	\end{equation}
	where we used $\phi_x' - \phi_x = \delta \phi_x$. The update rule again compensates the imaginary contributions in the transition probability.
	
	An update rule for the real part of the field can be derived in the same manner
	as for the Langevin equation. We complete the exponent of the product of the
	first Gaussian distribution and of the exponential function
	in Eq.~\eq{eq:ResultingTransitionComplexLangevin2} by
	\begin{equation}
	\frac{\epsilon}{2}\left[\frac{\delta
		S_\text{Re}}{\delta \phi_x} + \frac{\phi_x' -
		\phi_x}{2}\frac{\delta^2 S_\text{Re}}{\delta
		\phi_x^2}\right]\,.
	\end{equation}
	The argument of the Gaussian distribution in the transition
	probability~\eq{eq:TransitionProbabilityReal} now reads
	\begin{align}
	&\frac{\phi_x'-\phi_x}{\sqrt{2\epsilon}} +
	\sqrt{\frac{\epsilon}{2}}\left[\frac{\delta S_{\text{Re}}}{\delta \phi_x} + \frac{\phi_x' -
		\phi_x}{2}\frac{\delta^2 S_\text{Re}}{\delta
		\phi_x^2}\right]\nonumber\\[1ex]&\quad\quad\quad=\frac{\phi_x'-\phi_x}{\sqrt{2\epsilon}}\left[	1 + \frac{\epsilon}{2}\frac{\delta^2 S_\text{Re}}{\delta \phi_x^2}\right] +
	\sqrt{\frac{\epsilon}{2}}\frac{\delta S_{\text{Re}}}{\delta \phi_x}\,.
	\end{align}
	Because of the additional second order term, the normalization factor of the transition probability needs to be adjusted by the factor
	\begin{equation}
	1 + \frac{\epsilon}{2}\frac{\delta^2
		S_\text{Re}}{\delta \phi_x^2}\,.
	\end{equation}
	An explicit update rule can be
	derived again by performing a transformation of the probability density, cf.
	Eq.~\eq{eq:TransformationTransitionProbability} for more details. We finally arrive at
	\begin{equation}
	\phi_x' = \phi_x - \left(\epsilon \frac{\delta
		S_\text{Re}}{\delta \phi_x} + \sqrt{2\epsilon}
	\eta\right)\bigg/\left(1 + \frac{\epsilon}{2}
	\frac{\delta^2 S_\text{Re}}{\delta
		\phi_x^2}\right)\,.
	\end{equation}
	The update rule of the imaginary part now depends on the outcome of the real
	part. Accordingly, one has to update at first $\phi_x$ and then $\phi_y$.
	
	\subsection{Complex hat function algorithm}
	\label{sec:AppendixComplexHatFunctionAlgorithm}
	
	We derive a Langevin sampling by compensation algorithm for a different representation of the delta-distribution, namely, the triangular hat function. We consider again a complex action $S(\phi)$, as given, for example, in Eq.~\eq{eq:PolynomialModel} for the polynomial model. The derivation is in line with the systematic derivation in App.~\ref{sec:SystematicDerivation}.
	
	The hat function is given by
	\begin{equation}
	\label{eq:HatFunction}
	\eta_\epsilon(x) = \frac{1}{\epsilon} \begin{cases}
	1-\frac{x}{\epsilon}\quad&\textnormal{for}\;0\,\leq\, x \,<\, \epsilon\,,\\
	1+\frac{x}{\epsilon}\quad&\textnormal{for}\;-\epsilon\,<\, x \,<\,0\,,\\
	0\quad&\quad\text{otherwise.}
	\end{cases}
	\end{equation}
	We rewrite this, for simplicity, as
	\begin{equation}
	\eta_\epsilon(x) = \frac{1}{\epsilon} \left[1 - s \frac{x}{\epsilon}\right]\,\quad\text{for}\quad-\epsilon\,<\,x\,<\,\epsilon\,,
	\end{equation}
	where $s:=\text{sign}\left(x\right)$. Similar to the Gaussian distribution, the hat function converges, in the limit of $\epsilon\to 0$, to the delta-distribution.
	
	The corresponding transition probability is
	\begin{equation}
	\label{eq:HatTransitionProbability}
	W(\phi\to\phi')=\frac{1}{N}\left[1 - s \frac{\phi' - \phi}{\epsilon}\right] \times \exp\left(-\frac{\Delta S}{2}\right)\,,
	\end{equation}
	with $N$ being a normalization factor.

	With the same substitution as for complex Langevin dynamics,
	\begin{equation}
	\label{eq:ReplacingPhi}
	\phi\to\phi_x + \iu \phi_y\,,
	\end{equation}
	we identify the imaginary part $\phi_y$ as hidden dimension.
	
	After replacing $\phi$ by~\eq{eq:ReplacingPhi} in the transition probability~\eq{eq:HatTransitionProbability}, we identify
	\begin{align}
	\label{eq:HatFunctionProposalDistribution}
	q_s &= \frac{1}{\epsilon}\left[1 - \text{sign}_{\phi_x'-\phi_x} \frac{\phi_x' - \phi_x}{\epsilon}\right]\,,\nonumber\\[1ex]
	\bar{q}_s &= \frac{\iu}{\epsilon}\left[\text{sign}_{\phi_x'-\phi_x} \frac{\phi_y' - \phi_y}{\epsilon}\right]\,,
	\end{align}
	which are summed according to Eq.~\eq{eq:DecompositionProposalDistributionSum} and continue by decomposing the acceptance probability,
	\begin{align}
	\label{eq:HatFunctionAcceptanceProbability}
	A&(\phi', \phi) \propto \exp\left(-\frac{\Delta S_\text{Re}(\phi', \phi)}{2}\right) \nonumber\\[1ex] &\;\times \left[\cos\left(-\frac{\Delta S_\text{Im}(\phi', \phi)}{2}\right) + \iu \sin\left(-\frac{\Delta S_\text{Im}(\phi', \phi)}{2}\right)\right] \nonumber\\[1ex] &=\exp\left(-\frac{\Delta S_\text{Re}(\phi', \phi)}{2}\right) \times \left[ A_s + \bar{A}_s\right]\,,
	\end{align}
	with the action given by $S(\phi_x, \phi_y) = S_{\text{Re}}(\phi_x, \phi_y) + \iu S_{\text{Re}}(\phi_x, \phi_y)$. This allows distinguishing contributions that have no impact on detailed balance and contributions that need to vanish. The term $A_s$ is symmetric with respect to an exchange of $\phi_x'$ and $\phi_x$ whereas $\bar{A}_s$ is antisymmetric. Expanding the product of the proposal distribution and the acceptance probability gives a proposal distribution which is symmetric under an exchange of $\phi_x'$ and $\phi_x$, and which will drop out in the detailed-balance equation. Based on this, the transition probability for the real field $\phi_x$ is defined up to a normalisation factor as
	\begin{align}
	\label{eq:IntermediateHatFunctionTransitionProbability}
	T(&\phi_x'|\phi_x, \phi_y) \nonumber\\[1ex]
	&\propto \exp\left(-\frac{\Delta S_\text{Re}(\phi', \phi)}{2}\right)\times\left[q_s A_s + \bar{q}_s \bar{A}_s\right]\,.
	\end{align}
	The remaining terms must vanish,
	\begin{align}
	h&(\phi_y'|\phi_x', \phi_x, \phi_y) \nonumber\\[1ex]
	&=\exp\left(-\frac{\Delta S_\text{Re}(\phi', \phi)}{2}\right)\times\left[q_s \bar{A}_s + \bar{q}_s A_s\right] \stackrel{!}{=} 0\,,
	\end{align}
	for the not yet assigned parameter $\phi_y'$. This defines the update rule for $\phi_y$:
	\begin{equation}
	\label{eq:hatimag}
	\phi_y' = \phi_y+ \left[\frac{\epsilon}{s} - \left(\phi_x' - \phi_x\right)\right]\tan\left(-\frac{\Delta S_\text{Im}(\phi', \phi)}{2}\right)\,,
	\end{equation}
	and thus $g(\phi_y'|\phi_x', \phi_x, \phi_y)$. Because of the distinction of symmetric and antisymmetric parts, the transition probability $g$ posses Langevin symmetry
	\begin{equation}
	g(\phi_y'|\phi_x', \phi_x, \phi_y) = g(\phi_y'|\phi_x, \phi_x', \phi_y)\,.
	\end{equation}
	The update rule of the imaginary part can be used to further simplify the transition probability. We can solve the update rule~\eq{eq:hatimag} for $\bar{q}_s$ and insert the result into the transition probability to obtain
	\begin{equation}
	\label{eq:HatFunctionTransitionProbability}
	T(\phi_x'|\phi_x, \phi_y)\propto \exp\left(-\frac{\Delta S_\text{Re}(\phi', \phi)}{2}\right)\times q_s\left[A_s - \frac{\bar{A}_s^2}{A_s}\right]\,.
	\end{equation}
	The resulting transition probability $T$ leads to the same violation of the adapted detailed-balance equation~\eq{eq:DetailedBalanceLangevin} as for complex Langevin dynamics. The algorithm samples from the correct distribution only for $\epsilon\to 0$.
	
	\subsection{Uniform complex Langevin}
	
	Utilizing the results of the previous section, we define a sampling algorithm that uses a centred uniform distribution to propose states $\phi_x'$.
	
	This leads to the following ansatz for the transition probability $W(\phi\to\phi')$:
	\begin{align}
	\label{eq:UniformComplexLangevin}
	W(\phi& \to \phi')\nonumber\\[1ex]&\propto\int_{-l}^{l} \frac{\text{d}r}{2l}\,\delta\left(\phi' - (\phi + r)\right)\times \exp\left(-\frac{\Delta S(\phi', \phi)}{2}\right)\,,
	\end{align}
	where, in practice, $r$ is sampled from a uniform distribution in the interval $\left[-l, l\right]$. Replacing the delta-distribution by the triangular hat function~\eq{eq:HatFunction}, gives
	\begin{align}
	\label{eq:UniformComplexLangevinHF}
	W(\phi \to \phi')\propto\int_{-l}^{l} \frac{\text{d}r}{2l}\,\frac{1}{\epsilon}&\left[1 - \tilde{s}\frac{\phi' - (\phi + r)}{\epsilon}\right]\nonumber\\[1ex]& \times \exp\left(-\frac{\Delta S(\phi', \phi)}{2}\right)\,,
	\end{align}
	with
	\begin{equation}
	\tilde{s} = \text{sign}\left(\phi' - (\phi + r)\right)\,.
	\end{equation}
	Performing the same steps as for the complex hat function algorithm, this yields the update rule of the imaginary part,
	\begin{equation}
	\label{eq:UnformLangevinImag}
	\phi_y' = \phi_y+ \left[\frac{\epsilon}{\tilde{s}} - \left(\phi_x' - (\phi_x + r)\right)\right]\tan\left(-\frac{\Delta S_\text{Im}(\phi', \phi)}{2}\right)\,.
	\end{equation}
	To restore the original uniform distribution, we take the limit $\epsilon\to 0$. The update rule simplifies to
	\begin{equation}
	\label{eq:HatFunctionImagUpdateRule}
	\phi_y' = \phi_y + \left(\phi_x' - (\phi_x + r)\right)\tan\left(-\frac{\Delta S_\text{Im}(\phi', \phi)}{2}\right)\,.
	\end{equation}
	Following the same approach for the transition probability $T(\phi_x'|\phi_x, \phi_y)$, one arrives at
	\begin{align}
	\label{eq:UniformComplexLangevinRealAppendix}
	T&(\phi_x'|\phi_x, \phi_y)\propto\int_{-l}^{l} \frac{\text{d}r}{2l}\,\delta\left(\phi_x' - (\phi_x + r)\right)\nonumber\\[1ex]&\times\exp\left(-\frac{\Delta S_{\text{Re}}(\phi', \phi)}{2}\right)	
	\cos^{-1}\left(-\frac{\Delta S_\text{Im}(\phi', \phi)}{2}\right)\,.
	\end{align}
	According to the proposal distribution, the second term of the update rule~\eq{eq:HatFunctionImagUpdateRule} of the imaginary part vanishes, resulting in $\phi_y' = \phi_y$. To prevent this, we draw, in each update, two proposal states, $\phi_x'$ and $\tilde{\phi}_x'$, and adapt the update rule of the imaginary part:
	\begin{equation}
	\phi_y' = \phi_y + \left(\tilde{\phi}_x' - (\phi_x + r)\right)\tan\left(-\frac{\Delta S_\text{Im}(\phi', \phi)}{2}\right)\,.
	\end{equation}
	Similar to the complex hat function algorithm, the derived algorithm satisfies the constraints of the substitution algorithm only in the limit of infinitesimally small step sizes into the $\phi_x'$ direction. Based on the proposal distribution this can be implemented by considering the limit $l\to 0$.

	\section{Absorbing the imaginary contribution}
	\label{sec:AbsorbingTheImaginaryContribution}

	We start by considering
	\begin{align}
	\label{eq:TransitionComplexLangevinAppendix}
	T&(\phi_x|\phi_x', \phi_y) \exp\left(-\iu\Delta S_{\text{Im}}(\phi, \phi')\right)\propto\varphi\left(\frac{\phi_x-\phi_x'}{\sqrt{2\epsilon}}\right)\nonumber\\[1ex]&
	\times\exp\left(-\frac{\Delta S_{\text{Re}}(\phi, \phi')}{2}\right)\exp\left(-\iu\Delta S_{\text{Im}}(\phi, \phi')\right)\,.
	\end{align}
	Expanding $\Delta S_{\text{Re}}$ and $\Delta S_{\text{Im}}$ around $\phi_x'$, one obtains
	\begin{align}
	\label{eq:TransitionComplexLangevinAppendix2}
	&T(\phi_x|\phi_x', \phi_y)\exp\left(-\iu\Delta S_{\text{Im}}(\phi, \phi')\right)\nonumber\\[1ex]&\;\propto\exp\bigg(-\frac{1}{2}\left(\frac{\phi_x-\phi_x'}{\sqrt{2\epsilon}}\right)^2 \nonumber\\[1ex]&
	 -\frac{\phi_x - \phi_x'}{2}\bigg(\frac{\delta S_\text{Re}(\phi_x' + \iu \phi_y)}{\partial \phi_x'} +  2\iu\frac{\delta S_\text{Im}(\phi_x' + \iu \phi_y)}{\partial \phi_x'}\bigg)\bigg)\,.
	\end{align}	
	Next, we complete the square in the exponent,
	\begin{align}
	\label{eq:TransitionComplexLangevinAppendix3}
	T&(\phi_x|\phi_x', \phi_y)\exp\left(-\iu\Delta S_{\text{Im}}(\phi, \phi')\right)\propto\varphi\bigg(\frac{\phi_x-\phi_x'}{\sqrt{2\epsilon}}\, \nonumber\\[1ex]&
	\, + \sqrt{\frac{\epsilon}{2}}\left(\frac{\delta S_\text{Re}(\phi_x' + \iu \phi_y)}{\partial \phi_x'} + 2\iu\frac{\delta S_\text{Im}(\phi_x' + \iu \phi_y)}{\partial \phi_x'}\right)\bigg)\,.
	\end{align}
	We insert this expression into constraint~\eq{eq:LangevinConstraint}. As a result, an integration over $\phi_x$ is possible since the dependence on $\phi_x$ in the action was eliminated by the expansion around $\phi_x'$.
	
	\bibliography{literature}
	
\end{document}